\documentclass[a4paper,11pt]{article}
\pdfoutput=1
\usepackage{jheppub}

\usepackage{bm}
\usepackage{multirow}
\usepackage{comment}

\allowdisplaybreaks
\def\Babar{{\mbox{\slshape B\kern-0.1em{\smaller A}\kern-0.1em B\kern-0.1em{\smaller A\kern-0.2em R}}}}

\newcommand{\ts}{\hskip 0.08em}
\newcommand{\tts}{\hskip 0.04em}

\newcommand{\Kzerobar}{\overline{K}^0}
\newcommand{\Kstarzerobar}{\overline{K}^{\ts *0}}

\newcommand{\Bzerobar}{\overline{B}^0}
\newcommand{\Bsbar}{\overline{B}_s}

\newcommand{\Utt}{\mathcal{U}^{\ts\prime}_{\ts{\bm{\bar{3}}}{\bm{\bar{3}}}}}
\newcommand{\Uttp}{\mathcal{U}^{\ts\prime}_{\ts{\bm{\bar{3}}}{\bm{\bar{3}'}}}}

\newcommand{\Utptp}{\mathcal{U}^{\ts\prime}_{\ts{\bm{\bar{3}'}}{\bm{\bar{3}'}}}}

\title{\boldmath Revisiting rescattering contributions to $\overline{B}_{(s)}\to D^{(*)}_{(s)} M$ decays}

\author[a,b,c]{Motoi Endo,}
\author[c]{Syuhei Iguro,}
\author[a]{and Satoshi Mishima}

\affiliation[a]{Theory Center, IPNS, KEK, Tsukuba 305-0801, Japan}
\affiliation[b]{The Graduate University of Advanced Studies (Sokendai), Tsukuba 305-0801, Japan}
\affiliation[c]{Kavli Institute for the Physics and Mathematics of the Universe, University of Tokyo, Kashiwa 277-8583, Japan}

\emailAdd{motoi.endo@kek.jp}
\emailAdd{syuhei.iguro@ipmu.jp}
\emailAdd{satoshi.mishima@kek.jp}

\abstract{Motivated by the reported discrepancies between experimental data and Standard Model predictions for the branching ratios of the color-allowed decays $\overline{B}^0 \to D^{(*)+}K^-$ and $\overline{B}^0_s \to D^{(*)+}_s \pi^-$, we study final-state rescattering effects on $\overline{B}_{(s)}\to D^{(*)}_{(s)} P$ and $\overline{B}_{(s)}\to D_{(s)} V$, where $P$ ($V$) is a light pseudoscalar (vector) meson. 
We consider quasi-elastic rescatterings in the framework of $SU(3)$ and $U(3)$ flavor symmetries, and find that the effects cannot explain the measured branching ratios of the color-allowed and color-suppressed decays simultaneously. 
We also perform global fits to the $\overline{B}_{(s)}\to D^{(*)}_{(s)} P$ and $\overline{B}_{(s)}\to D_{(s)} V$ data, allowing for new physics contributions to the Wilson coefficient $a_1$ associated with the color-allowed tree amplitudes. 
It is shown that the fits prefer a downward shift of $\mathcal{O}(10\%)$ in $a_1$ even in the presence of the quasi-elastic rescattering contributions. 
}

\begin{document}
\preprint{IPMU21--0057, KEK--TH--2348}
\maketitle
\flushbottom

\section{Introduction}

Two-body hadronic $B$ decays into a heavy-light final state
$\overline{B}_{(s)}\to D_{(s)}^{(*)} M$, where $M$ is a light meson,
have played an important role in testing the factorization
hypothesis~\cite{Bjorken:1988kk}. 
In $\overline{B}_{(s)}^0\to D_{(s)}^+\pi^-$ and 
$\overline{B}_{(s)}^0\to D_{(s)}^+ K^-$ decays, which are dominated by
the so-called color-allowed tree amplitude, 
the hypothesis is expected to work well, 
since contributions from a gluon exchange between the light meson
emitted from the weak vertex and the soft spectator quark in a heavy
meson are power suppressed in the heavy-quark limit. Moreover, they 
have no penguin contribution.
For these decays the factorization theorem was proved 
at the leading power in the heavy-quark expansion with the QCD
factorization (QCDF) approach~\cite{Beneke:2000ry} 
and with the soft-collinear effective theory (SCET)~\cite{Bauer:2001cu}.
There are great improvements in theoretical calculations of the
color-allowed tree amplitude in recent years, which enable us to study
possible contributions to these decays from new-physics (NP) beyond the Standard Model (SM).

The branching ratios of 
$\overline{B}_{(s)}^0\to D_{(s)}^{(*)+}\pi^-$ and 
$\overline{B}_{(s)}^0\to D_{(s)}^{(*)+} K^-$ 
are calculated in QCDF to next-to-next-to-leading order (NNLO) 
accuracy~\cite{Huber:2016xod}. They have recently been updated 
in ref.~\cite{Bordone:2020gao} with new determinations 
of $\overline{B}_{(s)}\to D_{(s)}^{(*)}$ form factors 
in the heavy-quark expansion up to
$\mathcal{O}(\Lambda^2_{\mathrm{QCD}}/m_c^2)$~\cite{Bordone:2019guc}. 
The updated predictions for the branching ratios show a tension with
the experimental measurements with a significance of 
the $4.4\ts\sigma$ level~\cite{Bordone:2020gao}.\footnote{Similar results are obtained in ref.~\cite{Cai:2021mlt} with different inputs.}

The tension requires a downward shift of 10\% in the $b\to c\bar{u}q$ amplitude 
with $q$ being $d$ or $s$, which might be caused by NP.\footnote{
Implication of CP-violating NP phases is studied in ref.~\cite{Fleischer:2021cct}. }
It is shown in ref.~\cite{Iguro:2020ndk} that 
the tension can be reduced partly by introducing a new vector boson $W'$ 
without conflicting with narrow resonance searches at the Large Hadron Collider (LHC).\footnote{A model-independent NP analysis has also been carried out in ref.~\cite{Cai:2021mlt}. }

On the other hand, the experimental data for the branching ratios of 
$\overline{B}^0\to D^{(*)+}\pi^-$,
$\overline{B}^0\to D^{(*)0}\pi^0$ and 
$B^-\to D^{(*)0}\pi^-$ indicate a failure of the naive 
factorization~\cite{Cheng:2001sc,Neubert:2001sj}.\footnote{A different
 conclusion was obtained in refs.~\cite{Xing:2001nj,Xing:2003fe}.} 
Here $\overline{B}^0\to D^{(*)0}\pi^0$ is dominated by the
color-suppressed tree amplitude, while $B^-\to D^{(*)0}\pi^-$ has both the color-allowed and color-suppressed tree amplitudes.
For $a_1$ and $a_2$ being the Wilson coefficients associated with 
the color-allowed and color-suppressed tree amplitudes, respectively, 
a fit to the data prefers a larger $|a_2|$ compared to those extracted
from $\overline{B}\to J/\psi \overline{K}^{(*)}$ and  
$\overline{B}\to \psi(2S) \overline{K}^{(*)}$  
with a sizable relative phase between $a_1$ and $a_2$.
Similar results can also be obtained from other $\overline{B}_{(s)}\to
D_{(s)}^{(*)} M$
processes~\cite{Chiang:2002tv,Kim:2004hx,Chiang:2007bd}.
The relative phase between $a_1$
and $a_2$ implies that final-state interactions (FSI) are significant
in the color-suppressed modes.\footnote{In the PQCD approach based on
the $k_T$ factorization, the larger $|a_2|$ with the sizable phase
is explained by spectator-scattering contributions~\cite{Keum:2003js,Li:2008ts}.}

It is however difficult to calculate the FSI effects in 
$\overline{B}_{(s)}\to D_{(s)}^{(*)} M$ from first principles.\footnote{The FSI effects in heavy meson decays were explored first for the charmed meson decays in ref.~\cite{Blok:1986sn} using QCD sum rules.}. 
In SCET, the FSI contributions in the color-suppressed decays are 
factorized into soft factors~\cite{Mantry:2003uz,Blechman:2004vc},
which have to be calculated in non-perturbative methods or extracted
from data. Non-factorizable contributions from a single soft-gluon
exchange are estimated with the light-cone QCD sum rules (LCSR) 
for the color-allowed decays~\cite{Bordone:2020gao} 
and the color-suppressed decays~\cite{Halperin:1994hg,Cui:2004jc},
and the size of them for the former (latter) is found to be
insignificant (significant) compared to the factorizable ones. 
Also the rescattering of the final-state mesons are studied with various models~\cite{Donoghue:1996hz,Blok:1996uf,Suzuki:1999uc,Chua:2001br,Fayyazuddin:2002aa,Wolfenstein:2003pc,Calderon:2003vc,Fayyazuddin:2004ac,Wolfenstein:2004km,Cheng:2004ru,Chua:2005dt,Chua:2007qw,Suzuki:2007je}.

It is pointed out in ref.~\cite{Donoghue:1996hz} using the Regge approach that the final-state rescattering effects do not vanish in the large $b$-quark mass limit. 
On the other hand, in the QCDF approach, the rescattering effects are found to be suppressed in the heavy-quark limit~\cite{Beneke:2000ry}. 
Hence the suppression could be understood by a systematic cancellation among all intermediate hadronic states~\cite{Suzuki:1999uc,Beneke:2000ry}. 
Nonetheless, the rescattering effects may remain sizable, since the $b$-quark mass is not infinite in reality.

In this paper we study the {\it quasi-elastic} rescattering contributions to the $\overline{B}_{(s)}\to D_{(s)}^{(*)} M$ decays, following the framework developed in refs.~\cite{Chua:2001br,Chua:2005dt,Chua:2007qw}. 
The elastic $D\pi\to D\pi$ rescattering picture is extended to quasi-elastic $D^{(*)}P \to D^{(*)}P$ and $DV\to DV$ rescatterings by based on $SU(3)$ and $U(3)$ flavor symmetries, 
where $P$ and $V$ represent the pseudoscalar $SU(3)$ octet and the vector $U(3)$ nonet, respectively. 
We assume that inelastic-rescattering contributions are suppressed, and examine the question whether the quasi-elastic rescattering contributions can resolve the tensions between the theoretical predictions and the experimental data for the color-allowed decays. 
The analyses in refs.~\cite{Chua:2001br,Chua:2005dt,Chua:2007qw} are updated with latest theoretical and experimental information. 
In particular, the Wilson coefficient $a_1$ at NNLO~\cite{Huber:2016xod} and the $\overline{B}_{(s)}\to D_{(s)}^{(*)}$ form factors at  
$\mathcal{O}(\Lambda^2_{\mathrm{QCD}}/m_c^2)$~\cite{Iguro:2020cpg} are employed to make the calculation of the color-allowed tree amplitude more reliable.\footnote{It is assumed that the $SU(3)$ flavor symmetry holds between the $\overline{B}\to D^{(*)}$ and $\overline{B}_{s}\to D_{s}^{(*)}$ form factors as suggested in the result of refs.~\cite{Kobach:2019kfb,Bordone:2019guc}.}

This paper is organized as follows. 
In section~\ref{Sec:theory} the theoretical framework for calculating decay amplitudes is described under the presence of the quasi-elastic rescattering. 
In section~\ref{Sec:analysis} numerical analysis of the rescattering effects in the color-allowed decays is presented. 
Finally, the section~\ref{Sec:summary} is devoted to conclusions and discussion. 
Besides, appendix~\ref{Sec:fitresults} contains some details of numerical results.

\section{Theoretical framework}
\label{Sec:theory}

In this section the theoretical framework for calculating decay amplitudes is briefly introduced with quasi-elastic rescattering contributions. 
The decays of $\overline{B}\to DP$, $\overline{B}\to D^*P$, and $\overline{B}\to DV$ with $b\to c\bar{u}q$ transitions are studied, where $q$ is $d,s$, and $P$ ($V$) stands for a light pseudoscalar (vector) meson.\footnote{We do not consider $\overline{B}\to D^* V$ in this paper, since experimental uncertainties are large \cite{PDG:2020}.}

\subsection{Decay amplitudes}
\label{Sec:decay}

The weak effective Hamiltonian relevant for the current study consists only of the following tree operators:
\begin{align}
\mathcal{H}_W
=
\frac{4\ts G_F}{\sqrt{2}}\,
\sum_{q=d,s}
V_{cb} V_{uq}^*
\big( C_1 \mathcal{O}_1^{\tts q} + C_2 \mathcal{O}_2^{\tts q} \big)
+ \mathrm{h.c.},
\label{eq:Hamiltonian}
\end{align}
where $V_{ij}$ are the Cabibbo-Kobayashi-Maskawa (CKM) matrix elements~\cite{Cabibbo:1963yz,Kobayashi:1973fv}. 
The local operators are defined in the Chetyrkin-Misiak-M\"{u}nz (CMM) basis~\cite{Chetyrkin:1996vx,Chetyrkin:1997gb} as 
\begin{align}
\mathcal{O}_1^{\tts q}
&=
\big(\bar{c}\ts \gamma^{\mu} T^a P_L\tts b\big)
\big(\bar{q}\tts \gamma_{\mu} T^a P_L u\big)
\,,\\
\mathcal{O}_2^{\tts q}
&=
\big(\bar{c}\ts \gamma^{\mu} P_L\tts b\big)
\big(\bar{q} \tts\gamma_{\mu} P_L u\big)
\,,
\label{operator}
\end{align}
where $P_L=(1-\gamma_5)/2$, and $T^a$ is a generator of $SU(3)$. 
Suppose that there is no CP-violating phase in $\mathcal{H}_W$\footnote{Note that the CP-violating phase in the CKM matrix is irrelevant in the current study.}, 
the time-reversal invariance of $\mathcal{H}_W$ leads to
\begin{align}
\langle i; \mathrm{out} | \mathcal{H}_W | \overline{B} \rangle^*
= 
\sum_j S_{ji}^*\ts 
\langle j; \mathrm{out} | \mathcal{H}_W | \overline{B} \rangle 
\tts,
\end{align}
where the strong scattering $S$-matrix element $S_{ji}$ is defined in terms of {\it in} and {\it out} states as 
$S_{ji} = \langle j; \mathrm{out}|i; \mathrm{in} \rangle$. 
The solution of this relation is given by
$\langle i; \mathrm{out} | \mathcal{H}_W | \overline{B} \rangle
= \sum_j S_{ij}^{1/2} \mathcal{A}_j^{\tts 0}$,
where $\mathcal{A}_j^{\tts 0}$ is real~\cite{Suzuki:1999uc}.
A decay amplitude with FSI, denoted as 
$\mathcal{A}_i^{\tts\mathrm{FSI}}
=
\langle i; \mathrm{out} | \mathcal{H}_W | \overline{B} \rangle$, 
is given by 
\begin{align}
\mathcal{A}_i^{\tts\mathrm{FSI}}
=
\sum_j
S_{ij}^{1/2} 
\mathcal{A}_j^{\tts 0}\tts,
\end{align}
where $A_j^0$ represents the corresponding decay amplitude without FSI. 
Following the procedure described in ref.~\cite{Chua:2018ikx}, the scattering matrix is decomposed into a product of two matrices, 
$S_{ij}^{1/2} = \sum_k (S_{\mathrm{res}}^{1/2})_{ik} (S_{2}^{1/2})_{kj}$.
Then the amplitude is written as
\begin{align}
\mathcal{A}_i^{\tts\mathrm{FSI}}
=
\sum_k
(S_{\mathrm{res}}^{1/2})_{ik}\ts
\mathcal{A}_k^{\ts\mathrm{fact}}\tts,
\qquad 
\mathcal{A}_k^{\ts\mathrm{fact}}
=
\sum_j
(S_{2}^{1/2})_{kj}\ts
\mathcal{A}_j^{\tts 0}\tts,
\label{eq:FSI}
\end{align}
where $S_{\mathrm{res}}^{1/2}$ is a rescattering matrix, and $\mathcal{A}_k^{\ts\mathrm{fact}}$ is a decay amplitude which incorporates a part of FSI effects. For the color-allowed tree amplitude, $\mathcal{A}_k^{\ts\mathrm{fact}}$ is assumed to correspond to the amplitude calculated in the QCDF approach.

\begin{figure}[t]
\centering
\includegraphics[scale=1.0]{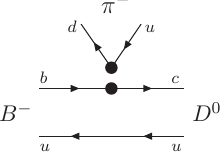}
\hspace{5mm}
\includegraphics[scale=1.0]{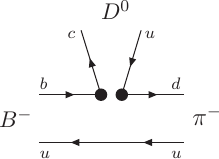}
\hspace{5mm}
\includegraphics[scale=1.0]{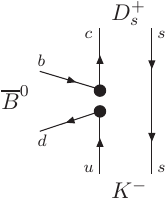}
\caption{Representative diagrams of the $b\to c\bar{u}d$ decays:
  the color-allowed tree diagram $T$ (left),
  the color-suppressed tree diagram $C$ (center),
  and the $W$-exchange diagram $E$ (right).
  A pair of blobs denotes a four-quark operator originating through 
  the weak interaction.}
\label{fig:diagrams}
\end{figure}

The decay amplitude $\mathcal{A}_k^{\ts\mathrm{fact}}$ can be expressed with topological amplitudes in the flavor-diagram approach based on the $SU(3)$ flavor symmetry~\cite{Zeppenfeld:1980ex,Savage:1989ub,Gronau:1995hm,Chiang:2002tv,Kim:2004hx,Colangelo:2005hh,Chiang:2007bd,Fleischer:2010ca}. 
There are three types of amplitudes for $\overline{B}_{(s)}\to D_{(s)}^{(*)} M$ decays:
the color-allowed tree amplitude $T$, the color-suppressed tree amplitude $C$,
and the $W$-exchange amplitude $E$, as shown in figure~\ref{fig:diagrams}. 
For example, the decay amplitude for $B^- \to D^0 \pi^-$ is written as
\begin{align}
\mathcal{A}(B^- \to D^0 \pi^-)
=
T_D+C_P\tts, 
\end{align}
where the subscript represents the final-state meson that involves the spectator quark.
The topological amplitudes for other channels 
are summarized in tables~\ref{tab:BDP}, \ref{tab:BDstP} and \ref{tab:BDV}, where $s_\phi$ and $c_\phi$ are the sine and cosine of the mixing angle $\phi_\eta$ in the $\eta$-$\eta'$ system.
The definition of the mixing angle is based on ref.~\cite{Escribano:2015yup}, and the single angle hypothesis $\phi_\eta\equiv \phi_q = \phi_s$ is adopted.

\begin{table}[t]
\centering
\begin{tabular}{cclcc}
\hline
Transition & $\{S,I_z\}$
&\hspace{4mm}Mode & Amplitude & Data
\\
\hline
$b\to c\bar{u}d$ 
& $\{1,-1\}$
& $\Bsbar \to D_s^+ \pi^-$
& $T_D$
& $30.0\pm 2.3$
\\
&
& $\Bsbar \to D^0 K^0$
& $C_P$
& $4.3\pm 0.9$
\\
\cline{2-5}
& $\{0,-3/2\}$
& $B^- \to D^0 \pi^-$
& $T_D+C_P$ 
& $46.8\pm 1.3$
\\
\cline{2-5}
& $\{0,-1/2\}$
& $\Bzerobar \to D^+ \pi^-$
& $T_D+E$ 
& $25.2\pm 1.3$
\\
&
& $\Bzerobar \to D^0 \pi^0$
& $\frac{1}{\sqrt{2}}(-C_P+E)$ 
& $2.63\pm 0.14$
\\  
&
& $\Bzerobar \to D^0 \eta$
& $\frac{c_\phi}{\sqrt{2}}(C_P+E)$
& $2.36\pm 0.32$
\\
&
& $\Bzerobar \to D^0 \eta'$
& $\frac{s_\phi}{\sqrt{2}}(C_P+E)$
& $1.38\pm 0.16$
\\
&
& $\Bzerobar \to D_s^+ K^-$
& $E$
& $0.27\pm 0.05$
\\
\hline
$b\to c\bar{u}s$ 
& $\{-1,0\}$
& $\Bzerobar \to D^+ K^-$
& $T_D$ 
& $1.86\pm 0.20$
\\
&
& $\Bzerobar \to D^0 \Kzerobar$
& $C_P$ 
& $0.52\pm 0.07$
\\
\cline{2-5}
& $\{-1,-1\}$
& $B^- \to D^0 K^-$
& $T_D+C_P$ 
& $3.63\pm 0.12$
\\
\cline{2-5}
& $\{0,-1/2\}$
& $\Bsbar \to D_s^+ K^-$
& $T_D+E$ 
& $2.27\pm 0.19\, {}^{\dagger}$
\\
&
& $\Bsbar \to D^0 \eta$
& $-s_\phi\,C_P + \frac{c_\phi}{\sqrt{2}}\ts E$ 
& ---
\\
&
& $\Bsbar \to D^0 \eta'$
& $c_\phi\,C_P + \frac{s_\phi}{\sqrt{2}}\ts E$ 
& ---
\\
&
& $\Bsbar \to D^0 \pi^0$
& $\frac{1}{\sqrt{2}}E$ 
& ---
\\
&
& $\Bsbar \to D^+ \pi^-$
& $E$ 
& ---
\\
\hline
\end{tabular}
\caption{List of $\overline{B}\to DP$ decays with 
$b\to c\bar{u}d$ and $b\to c\bar{u}s$ transitions. 
The values in the second column are the strangeness and the $z$ component of isospin of the final state.
The fourth column shows the $SU(3)$ decomposition of the decay amplitudes.  
The experimental data for the CP-averaged branching ratios (in units of $10^{-4}$) in the fifth column correspond to the PDG values~\cite{PDG:2020}, which are consistent with the HFLAV ones~\cite{HFLAV:2019otj}. 
${}^{\dagger}$The data for $\Bsbar \to D_s^+ K^-$
represents a sum of CP-averaged branching ratios,
$\mathcal{B}(\Bsbar \to D_s^+ K^-)$ and 
$\mathcal{B}(\Bsbar \to D_s^- K^+)$. 
}
\label{tab:BDP}
\end{table}
\begin{table}[t]
\centering
\begin{tabular}{cclcc}
\hline
Transition & $\{S,I_z\}$
&\hspace{4mm}Mode & Amplitude & Data
\\
\hline
$b\to c\bar{u}d$ 
& $\{1,-1\}$
& $\Bsbar \to D_s^{*+} \pi^-$
& $T_D$
& $20\pm 5$
\\
&
& $\Bsbar \to D^{*0} K^0$
& $C_P$
& $2.8\pm 1.1$
\\
\cline{2-5}
& $\{0,-3/2\}$
& $B^- \to D^{*0} \pi^-$
& $T_{D^*}+C_P$ 
& $49.0\pm 1.7$
\\
\cline{2-5}
& $\{0,-1/2\}$
& $\Bzerobar \to D^{*+} \pi^-$
& $T_{D^*}+E$ 
& $27.4\pm 1.3$
\\
&
& $\Bzerobar \to D^{*0} \pi^0$
& $\frac{1}{\sqrt{2}}(-C_P+E)$ 
& $2.2\pm 0.6$
\\
&
& $\Bzerobar \to D^{*0} \eta$
& $\frac{c_\phi}{\sqrt{2}}(C_P+E)$
& $2.3\pm 0.6$
\\
&
& $\Bzerobar \to D^{*0} \eta'$
& $\frac{s_\phi}{\sqrt{2}}(C_P+E)$
& $1.40\pm 0.22$
\\
&
& $\Bzerobar \to D_s^{*+} K^-$
& $E$
& $0.22\pm 0.03$
\\
\hline
$b\to c\bar{u}s$ 
& $\{-1,0\}$
& $\Bzerobar \to D^{*+} K^-$
& $T_{D^*}$ 
& $2.12\pm 0.15$
\\
&
& $\Bzerobar \to D^{*0} \Kzerobar$
& $C_P$ 
& $0.36\pm 0.12$
\\
\cline{2-5}
& $\{-1,-1\}$
& $B^- \to D^{*0} K^-$
& $T_{D^*}+C_P$ 
& $3.97^{+0.31}_{-0.28}$
\\
\cline{2-5}
& $\{0,-1/2\}$
& $\Bsbar \to D_s^{*+} K^-$
& $T_{D^*}+E$ 
& $1.33\pm 0.35\, {}^{\dagger}$
\\
&
& $\Bsbar \to D^{*0} \eta$
& $-s_\phi\,C_P + \frac{c_\phi}{\sqrt{2}}\ts E$ 
& ---
\\
&
& $\Bsbar \to D^{*0} \eta'$
& $c_\phi\,C_P + \frac{s_\phi}{\sqrt{2}}\ts E$ 
& ---
\\
&
& $\Bsbar \to D^{*0} \pi^0$
& $\frac{1}{\sqrt{2}}E$ 
& ---
\\
&
& $\Bsbar \to D^{*+} \pi^-$
& $E$ 
& ---
\\
\hline
\end{tabular}
\caption{List of $\overline{B}\to D^*P$ decays with 
$b\to c\bar{u}d$ and $b\to c\bar{u}s$ transitions. 
See also the caption of table~\ref{tab:BDP}. 
${}^{\dagger}$The data for $\Bsbar \to D_s^{*+} K^-$
represents a sum of CP-averaged branching ratios
$\mathcal{B}(\Bsbar \to D_s^{*+} K^-)$ and 
$\mathcal{B}(\Bsbar \to D_s^{*-} K^+)$. 
}
\label{tab:BDstP}
\end{table}
\begin{table}[t]
\centering
\begin{tabular}{cclcc}
\hline
Transition & $\{S,I_z\}$
&\hspace{4mm}Mode & Amplitude & Data
\\
\hline
$b\to c\bar{u}d$ 
& $\{1,-1\}$
& $\Bsbar \to D_s^+ \rho^-$
& $T_D$
& $69\pm 14$
\\
&
& $\Bsbar \to D^0 K^{*0}$
& $C_V$
& $4.4\pm 0.6$
\\
\cline{2-5}
& $\{0,-3/2\}$
& $B^- \to D^0 \rho^-$
& $T_D+C_V$ 
& $134\pm 18$
\\
\cline{2-5}
& $\{0,-1/2\}$
& $\Bzerobar \to D^+ \rho^-$
& $T_D+E$ 
& $76\pm 12$
\\
&
& $\Bzerobar \to D^0 \rho^0$
& $\frac{1}{\sqrt{2}}(-C_V+E)$ 
& $3.21\pm 0.21$
\\
&
& $\Bzerobar \to D^0 \omega$
& $\frac{1}{\sqrt{2}}(C_V+E)$ 
& $2.54\pm 0.16$
\\
& 
& $\Bzerobar \to D^0  \phi$
& $0$
& ---
\\
&
& $\Bzerobar \to D_s^+ K^{*-}$
& $E$
& $0.35\pm 0.10$
\\
\hline
$b\to c\bar{u}s$ 
& $\{-1,0\}$
& $\Bzerobar \to D^+ K^{*-}$
& $T_D$ 
& $4.5\pm 0.7$
\\
&
& $\Bzerobar \to D^0 \Kstarzerobar$
& $C_V$ 
& $0.45\pm 0.06$
\\
\cline{2-5}
& $\{-1,-1\}$
& $B^- \to D^0 K^{*-}$
& $T_D+C_V$ 
& $5.3\pm 0.4$ 
\\
\cline{2-5}
& $\{0,-1/2\}$
& $\Bsbar \to D_s^+ K^{*-}$
& $T_D+E$ 
& ---
\\
&
& $\Bsbar \to D^0 \phi$
& $-C_V$ 
& $0.30\pm 0.05$
\\
&
& $\Bsbar \to D^0 \omega$
& $\frac{1}{\sqrt{2}}E$ 
& ---
\\
&
& $\Bsbar \to D^0 \rho^0$
& $\frac{1}{\sqrt{2}}E$ 
& ---
\\
&
& $\Bsbar \to D^+ \rho^-$
& $E$ 
& ---
\\
\hline
\end{tabular}
\caption{List of $\overline{B}\to DV$ decays with 
$b\to c\bar{u}d$ and $b\to c\bar{u}s$ transitions. 
See also the caption of table~\ref{tab:BDP}.}
\label{tab:BDV}
\end{table}

In the factorization approach, 
the amplitudes $T_D$ and $C_P$ are given by 
\begin{align}
T_D
&=
\frac{G_F}{\sqrt{2}}\ts 
a_1(\mu)\ts
(m_B^2 - m_D^2)\ts
f_P\ts
f_0^{B\to D}(m_P^2)
\quad\qquad \mathrm{for}\ \overline{B}\to DP
\tts,
\label{eq:T_BDP}
\\
C_P
&=
\frac{G_F}{\sqrt{2}}\ts 
a_2(\mu)\ts
(m_B^2 - m_P^2)\ts
f_D\ts
f_0^{B\to P}(m_D^2)
\quad\qquad \mathrm{for}\ \overline{B}\to DP
\tts,
\label{eq:C_BDP}
\\
T_{D^*}
&=
\frac{G_F}{\sqrt{2}}\ts 
a_1(\mu)\,
2\ts m_{D^*} (\epsilon^*\cdot p_B)\ts
f_P\ts
A_0^{B\to D}(m_P^2)
\quad\quad \mathrm{for}\ \overline{B}\to D^*P
\tts,
\label{eq:T_BDstP}
\\
C_P
&=
\frac{G_F}{\sqrt{2}}\ts 
a_2(\mu)\,
2\ts m_{D^*} (\epsilon^*\cdot p_B)\ts
f_{D^*}\ts
f_+^{B\to P}(m_{D^*}^2)
\quad\, \mathrm{for}\ \overline{B}\to D^*P
\tts,
\label{eq:C_BDstP}
\\
T_D
&=
\frac{G_F}{\sqrt{2}}\ts 
a_1(\mu)\,
2\ts m_V (\epsilon^*\cdot p_B)\ts
f_V\ts
f_+^{B\to D}(m_V^2)
\qquad\,\, \mathrm{for}\ \overline{B}\to DV
\tts,
\label{eq:T_BDV}
\\
C_V
&=
\frac{G_F}{\sqrt{2}}\ts 
a_2(\mu)\,
2\ts m_V (\epsilon^*\cdot p_B)\ts
f_{D}\ts
A_0^{B\to V}(m_{D}^2)
\qquad\,\, \mathrm{for}\ \overline{B}\to DV
\tts,
\label{eq:C_BDV}
\end{align}
where $\epsilon^{*\mu}$ is the polarization vector of $D^*$ or $V$.
The definitions of 
the decay constants 
$f_P$, $f_V$ and $f_{D^{(*)}}$ and the transition form factors
$f_0^{B\to D}$,
$f_0^{B\to P}$, 
$f_+^{B\to P}$, 
$f_+^{B\to D}$,
$A_0^{B\to D}$ and 
$A_0^{B\to V}$ are found in ref.~\cite{Gubernari:2018wyi}. 
The coefficients $a_1(\mu)$ and $a_2(\mu)$ are evaluated at the
scale $\mu=m_b$, where the leading-order contributions are given by  
$a_1(\mu) = C_2(\mu)$ and $a_2(\mu) = 4\ts C_1(\mu)/9 + C_2(\mu)/3$
in terms of the Wilson coefficients in eq.~\eqref{eq:Hamiltonian}.

The coefficient $a_1(\mu)$ associated with the color-allowed tree amplitude is calculated for different processes in the framework of QCDF to the NNLO accuracy~\cite{Huber:2016xod}: 
\begin{align}
a_1(m_b)
=
\begin{cases}
\big(1.069^{+0.009}_{-0.012}\big)
+
\big(0.046^{+0.023}_{-0.015}\big)\ts i
&\mathrm{for}\  \overline{B}^0\to D^+ K^-,
\\
\big(1.072^{+0.011}_{-0.013}\big)
+
\big(0.044^{+0.022}_{-0.014}\big)\ts i
&\mathrm{for}\ \overline{B}^0\to D^+ \pi^-,
\\
\big(1.068^{+0.010}_{-0.012}\big)
+
\big(0.034^{+0.017}_{-0.011}\big)\ts i
&\mathrm{for}\ \overline{B}^0\to D^{*+} K^-,
\\
\big(1.071^{+0.012}_{-0.013}\big)
+
\big(0.032^{+0.016}_{-0.010}\big)\ts i
&\mathrm{for}\ \overline{B}^0\to D^{*+} \pi^-.
\end{cases}
\label{eq:a1}
\end{align}
It is noticed that these are almost universal, and hence a common value of $a_1(m_b)$ is used in our numerical analysis: 
\begin{align}
a_1(m_b) = 1.070 \pm 0.012\,. 
\label{eq:a1univ}
\end{align}
We set $a_1$ to be real by absorbing its phase in $a_2$ and $E$, which are complex parameters in the following analysis. 
On the other hand, it is hard to evaluate the color-suppressed tree amplitudes reliably in QCDF, because 
subleading-power and non-factorizable contributions are expected to be significant. 
In our analysis the effective coefficient $a_2^{\mathrm{eff}}$ is considered to include these contributions and taken to be a free parameter. 
Also, $SU(3)$ breaking effects in the amplitudes are simply assumed to be described by the factorization form as in eqs.~\eqref{eq:C_BDP}, \eqref{eq:C_BDstP} and \eqref{eq:C_BDV}. 
Furthermore, the $W$-exchange amplitude $E$, which is power suppressed compared to the color-allowed amplitude $T$, also cannot be calculated in the QCDF approach~\cite{Beneke:2000ry}. Therefore $E$ is taken as another free parameter under the assumption of the $SU(3)$ flavor symmetry.

Finally, the partial decay widths for the $b\to c\bar{u}q$ channels ($q=d,s$) are expressed with the decay amplitudes as 
\begin{align} 
\Gamma
=
\frac{p_{\mathrm{cm}}}{8\tts\pi\tts m_B^2}\ts
\big| V_{cb}V_{uq}^*\, \mathcal{A}\,\big|^2,
\end{align}
where $p_{\mathrm{cm}}$ is the magnitude of the three momentum of a final-state meson in the center-of-mass frame.

\subsection{Quasi-elastic rescattering}
\label{Sec:res}

The framework of the quasi-elastic rescattering in $\overline{B}_{(s)}\to D_{(s)}^{(*)} M$ was developed in refs.~\cite{Chua:2001br,Chua:2005dt,Chua:2007qw}. 
In the quasi-elastic picture, the rescattering occurs among different final states carrying the same quantum numbers.
Since QCD respects the $SU(3)$ flavor symmetry to some extent, the rescattering is considered among the states that reside in the same multiplet under the symmetry.
In the case of $\overline{B}\to DV$, the $U(3)$ symmetry instead of $SU(3)$ can be applied, since there is no $U(1)$ axial anomaly for the vector mesons.
Moreover, following the procedure in ref.~\cite{Chua:2018ikx}, we include a part of the $SU(3)$ breaking effects through the meson decay constants.

Under the $SU(3)$ flavor symmetry 
the $D^{(*)}P$ final states are decomposed into one $\bm{\overline{15}}$, one $\bm{6}$ and two $\bm{\overline{3}}$ representations, since 
the light pseudo-scalar mesons transform as ${\bm{8}}$ and ${\bm{1}}$, and the charm mesons $D^{(*)0}$, $D^{(*)+}$ and $D_s^{(*)+}$ form a $\bm{\overline{3}}$ representation.
Therefore, the rescattering matrix in eq.~\eqref{eq:FSI} can be written in the $SU(3)$ basis as~\cite{Chua:2001br}
\begin{align}
S_{\mathrm{res}}^{1/2}
=
e^{i\tts\delta_{\bm{\overline{15}}}}
\sum_{a=1}^{15}
\big| {\bm{\overline{15}}}; a \big\rangle
\big\langle {\bm{\overline{15}}}; a \big|
+
e^{i\tts\delta_{\bm{6}}}
\sum_{b=1}^{6}
\big| {\bm{6}}; b \big\rangle
\big\langle {\bm{6}}; b \big|
+
\sum_{m,n={\bm{\bar{3}}},{\bm{\bar{3}'}}}
\sum_{c=1}^{3}
\big| m; c \big\rangle\,
\mathcal{U}^{1/2}_{mn}
\,\big\langle n; c \big|
\tts,
\label{eq:S12}
\end{align}
where $\mathcal{U}^{1/2}$ is a $2\times 2$ symmetric unitary matrix.
Since the overall phase is not physical, the phase difference, 
\begin{align}
\delta' \equiv \delta_{\bm{6}} - \delta_{\bm{\overline{15}}}
\tts,
\label{eq:deltap}
\end{align}
is defined, and $\mathcal{U}^{1/2}$ is parameterized
in terms of three parameters $\theta$, $\sigma$ and $\tau$ (see, e.g., ref.~\cite{Suzuki:2007je}): 
\begin{align}
\mathcal{U}^{\ts\prime}_{mn}
\equiv
\mathcal{U}^{1/2}
e^{-i\tts\delta_{\bm{\overline{15}}}}
=
\left(\!
\begin{array}{cc}
\cos\tau & \sin\tau \\
-\sin\tau & \cos\tau
\end{array}
\!\right)
\left(\!
\begin{array}{cc}
e^{i\tts\theta} & 0 \\
0 & e^{i\tts\sigma}
\end{array}
\!\right)
\left(\!
\begin{array}{cc}
\cos\tau & -\sin\tau \\
\sin\tau & \cos\tau
\end{array}
\!\right).
\label{eq:Upmn}
\end{align}
Thus the rescattering matrix is parameterized by the four parameters $\delta'$, $\theta$, $\sigma$ and $\tau$. 
Note that the rescattering effects are turned off at $\delta'=\theta=\sigma=\tau=0$. 

The rescattering in the $D^{(*)}V$ final states can be parameterized in a similar way.
The light vector mesons form a $U(3)$ nonet instead of a $SU(3)$ octet. 
In this case the rescattering parameters have specific values as follows, which are called
``solution~1'' and ``solution~2'' in ref.~\cite{Chua:2001br}: 
\begin{align}
\bullet\ \mathrm{Solution}~1:&\quad
\delta'= \theta,\quad 
\sigma=0\tts,\quad
\sin\tau=\sqrt{\frac{1}{3}}\tts,\quad
\cos\tau=\sqrt{\frac{2}{3}}\tts,
\label{eq:U3sol1}
\\
\bullet\ \mathrm{Solution}~2:&\quad
\delta'= \theta = 0,\quad
\sin\tau=\frac{2\sqrt{2}}{3}\tts,\quad
\cos\tau=\frac{1}{3}\tts. 
\label{eq:U3sol2}
\end{align}
Hence only a single parameter is left free for each solution. 

In the following the rescattering formulae are presented for the $\overline{B}\to D P$ decays.
The corresponding formulae for $\overline{B}\to D^* P$ and $\overline{B}\to D V$ can be obtained by substituting the final-state mesons appropriately.

The final states are classified by the strangeness and the isospin. 
There are three sets for the decays with the $b\to c\bar{u}d$ transition: 
$\{S,I_z\}=\{0,-3/2\}$, $\{0,-1/2\}$ and $\{1,-1\}$. 
For $\overline{B}\to D P$, 
$\{S,I_z\}=\{0,-3/2\}$ corresponds only to the single process $\overline{B}\to D^{0} \pi^-$. 
Then the elastic rescattering causes only an unphysical overall phase shift,
and thus, does not alter the branching ratio of this channel. 
For the other sets the decay amplitudes are assembled as 
\begin{align}
\mathcal{A}_{d,0, -1/2}
=
\begin{pmatrix}
\mathcal{A}(\Bzerobar \to D^{+} \pi^-)
\\[1mm]
\mathcal{A}(\Bzerobar \to D^{0} \pi^0)
\\[1mm]
\mathcal{A}(\Bzerobar \to D_s^{+} K^-)
\\[1mm]
\mathcal{A}(\Bzerobar \to D^{0}  \eta_8)
\\[1mm]
\mathcal{A}(\Bzerobar \to D^{0}  \eta_1)
\end{pmatrix},
\qquad
\mathcal{A}_{d,1, -1}
=
\begin{pmatrix}
\mathcal{A}(\Bsbar \to D_s^{+} \pi^-)
\\[1mm]
\mathcal{A}(\Bsbar \to D^{0} K^0)
\end{pmatrix},
\end{align}
where the subscripts denote the flavor of the down-type quark, 
the strangeness and the isospin of the final states.  
The physical states, $\eta$ and $\eta'$, are given in terms of the octet state $\eta_8$ and
the singlet one $\eta_1$ as
$\eta = \eta_8\cos\theta_\eta - \eta_1\sin\theta_\eta$ and 
$\eta' = \eta_8\sin\theta_\eta + \eta_1\cos\theta_\eta$ 
with $\theta_\eta=\phi_\eta -\mathrm{arctan}\sqrt{2}$.

Similarly, the final states in the processes with the $b\to c\bar{u}s$ transition are classified into 
$\{S,I_z\}=\{-1,-1\}$, $\{-1,0\}$ and $\{0,-1/2\}$. For $\{S,I_z\}=\{-1,-1\}$, there is only the single process $B^-\to D^{0} K^-$. The decay amplitudes for the others are written as
\begin{align}
\mathcal{A}_{s,-1, 0}
=
\begin{pmatrix}
\mathcal{A}(\Bzerobar \to D^{+} K^-)
\\[1mm]
\mathcal{A}(\Bzerobar \to D^{0} \Kzerobar)
\end{pmatrix},
\qquad
\mathcal{A}_{s,0, -1/2}
=
\begin{pmatrix}
\mathcal{A}(\Bsbar \to D^{+} \pi^-)
\\[1mm]
\mathcal{A}(\Bsbar \to D^{0} \pi^0)
\\[1mm]
\mathcal{A}(\Bsbar \to D_s^{+} K^-)
\\[1mm]
\mathcal{A}(\Bsbar \to D^{0} \eta_{8})
\\[1mm]
\mathcal{A}(\Bsbar \to D^{0} \eta_{1})
\end{pmatrix}.
\end{align}

According to eq.~\eqref{eq:FSI}, the rescattering is incorporated into the decay amplitudes as 
\begin{align}
\mathcal{A}_{q, S, I_z}^{\tts\mathrm{FSI}}
=
V_{S, I_z}^{-1}\,
\mathcal{S}_{S, I_z}^{1/2}\,
V_{S, I_z}\,
\mathcal{A}_{q, S, I_z}^{\ts\mathrm{fact}}
\tts,
\label{eq:AmpFSI}
\end{align}
where $\mathcal{S}_{0,-1/2}^{1/2}$ is the rescattering matrix, and 
the matrix $V_{S,I_z}$ represents the $SU(3)$ breaking effects~\cite{Chua:2018ikx}. 
It is noticed that the breaking effects are removed from 
$\mathcal{A}_{q, S, I_z}^{\ts\mathrm{fact}}$ before the rescattering
by applying $V_{S, I_z}$, and then put them back after the rescattering by
$V_{S, I_z}^{-1}$~\cite{Chua:2018ikx}, since the rescattering is considered 
to work under the $SU(3)$ symmetry.

The explicit expressions of the symmetric matrices $\mathcal{S}_{S, I_z}^{1/2}$ 
are given by~\cite{Chua:2001br,Chua:2005dt,Chua:2007qw} 
\begin{align}
\mathcal{S}_{1,-1}^{1/2}
&=
\mathcal{S}_{-1,0}^{1/2}
=
\frac{e^{i\tts\delta_{\bm{\overline{15}}}}}{2}
\begin{pmatrix}
1 + e^{i\tts\delta'}
& 1 - e^{i\tts\delta'}
\\[1mm]
& 1 + e^{i\tts\delta'}
\end{pmatrix},
\label{eq:res2x2}
\\
\mathcal{S}_{0,-1/2}^{1/2} 
&=
\frac{e^{i\tts\delta_{\bm{\overline{15}}}}}{8}
\begin{pmatrix}
3 + u_+
& -\frac{1}{\sqrt{2}}\big( 5 - u_+ \big)
& -1 - u_-
& \sqrt{\frac{3}{2}}\big( 1 - v_-\big)
& \sqrt{6}\ts u'
\\[2mm]
& \frac{1}{2}\big( 11 + u_+ \big)
& -\frac{1}{\sqrt{2}}\big( 1 + u_- \big)
& \frac{\sqrt{3}}{2}\big( 1 - v_- \big)
& \sqrt{3}\ts u'
\\[2mm]
&
& 3 + u_+
& -\sqrt{\frac{3}{2}}\big( 3 - v_+ \big)
& \sqrt{6}\ts u'
\\[2mm]
&
&
& \frac{1}{2}\big( 9 + w_+ \big)
& u'
\\[2mm]
&
&
&
& u''
\end{pmatrix},
\end{align} 
where 
$u_\pm = 2 \ts e^{i\tts\delta'} \pm 3\, \Utt$, 
$v_\pm = 2 \ts e^{i\tts\delta'} \pm \Utt$, 
$w_+ = 6 \ts e^{i\tts\delta'} + \Utt$, 
$u' = 2\,\Uttp$ and 
$u''= 8\, \Utptp$, and the lower components are omitted for simplicity. 
The matrix $V_{-1,0}$ is a unit matrix,
and the others are modeled by the ratios of the meson decay constants (see ref.~\cite{Chua:2018ikx}): 
\begin{align}
V_{1,-1}
&=
\mathrm{diag}
\bigg(
1\tts,\
\frac{f_{D_s} f_\pi}{f_{D} f_K}
\bigg)
\tts,
\\
V_{0,-1/2}
&=
\mathrm{diag}
\bigg(
1\tts,\
1\tts,\
\frac{f_{D} f_\pi}{f_{D_s} f_K}\tts,\
\frac{f_{D} f_\pi}{f_{D} f_{\eta_8}}\tts,\
\frac{f_{D} f_\pi}{f_{D} f_{\eta_1}}
\bigg)
\tts.
\end{align}

For the $\overline{B}\to DV$ decays with $\{S,I_z\}=\{0,-1/2\}$, we calculate
the amplitudes $\mathcal{A}_{q, 0, -1/2}^{\ts\mathrm{fact}}$ in the basis with $\phi$ and $\omega$ instead of the $SU(3)$ states:
\begin{align}
\mathcal{A}_{d,0, -1/2}
=
\begin{pmatrix}
\mathcal{A}(\Bzerobar \to D^+ \rho^-)
\\[1mm]
\mathcal{A}(\Bzerobar \to D^0 \rho^0)
\\[1mm]
\mathcal{A}(\Bzerobar \to D_s^+ K^{*-})
\\[1mm]
\mathcal{A}(\Bzerobar \to D^0  \phi)
\\[1mm]
\mathcal{A}(\Bzerobar \to D^0  \omega)
\end{pmatrix},
\qquad
\mathcal{A}_{s,0, -1/2}
=
\begin{pmatrix}
\mathcal{A}(\Bsbar \to D^+ \rho^-)
\\[1mm]
\mathcal{A}(\Bsbar \to D^0 \rho^0)
\\[1mm]
\mathcal{A}(\Bsbar \to D_s^+ K^{*-})
\\[1mm]
\mathcal{A}(\Bsbar \to D^0 \phi)
\\[1mm]
\mathcal{A}(\Bsbar \to D^0 \omega)
\end{pmatrix}. 
\end{align}
In this case the rescattering formula includes the mixing of the states as 
\begin{align}
\mathcal{A}_{q, 0, -1/2}^{\tts\mathrm{FSI}}
&=
(V_{0,-1/2}^{DV})^{-1}\ts
(U^{DV})^{\dagger}\,
\mathcal{S}_{0,-1/2}^{1/2}\,
U^{DV}
V_{0,-1/2}^{DV}\,
\mathcal{A}_{q, 0, -1/2}^{\ts\mathrm{fact}}
\tts,
\end{align}
where $U^{DV}$ and $V_{2}^{DV}$ are defined by 
\begin{align}
U^{DV}
&=
\left(
\begin{array}{ccccc}
1 & 0 & 0 & 0 & 0
\\
0 & 1 & 0 & 0 & 0
\\
0 & 0 & 1 & 0 & 0
\\
0 & 0 & 0 &
\cos\theta_V & \sin\theta_V
\\
0 & 0 & 0 &
- \sin\theta_V & \cos\theta_V
\end{array}
\right),
\\[2mm]
V_{0,-1/2}^{DV}
&=
\mathrm{diag}
\bigg(
1\tts,\
1\tts,\
\frac{f_D f_\rho}{f_{D_s} f_{K^*}}\tts,\
\frac{f_D f_\rho}{f_{D} f_{\phi}}\tts,\
\frac{f_D f_\rho}{f_{D} f_{\omega}}
\bigg)
\tts.
\end{align}
The angle $\theta_V$ represents the mixing of the octet and singlet states: 
$\phi = \omega_8\cos\theta_V - \omega_1\sin\theta_V$ and 
$\omega = \omega_8\sin\theta_V + \omega_1\cos\theta_V$ with 
$\cos\theta_V = \sqrt{2/3}$ and
$\sin\theta_V = 1/\sqrt{3}$.

\section{Numerical analysis}
\label{Sec:analysis}

The input parameters for our numerical analysis are summarized in table~\ref{tab:parameters}. 
The decay constants and the mixing angle for the octet and singlet states of the pseudo-scalar mesons are chosen as $f_8 = 1.27\ts f_\pi$, 
$f_1 = 1.14\ts f_\pi$, and $\theta_\eta=\phi_\eta -\mathrm{arctan}\sqrt{2}$ with 
$\phi_{\eta} = 0.7036$ rad~\cite{Escribano:2015yup}. 
Other parameters, such as the Fermi constant $G_F$ and the masses and the lifetimes of the mesons, are taken from ref.~\cite{PDG:2020}.
Note that the value of $V_{cb}$ in this section corresponds to the so-called exclusive $V_{cb}$. If the inclusive $V_{cb}$, which is larger than the exclusive one by about 5$\%$~\cite{PDG:2020}, is used, the tensions between the experimental data and the theoretical predictions for the color-allowed decays are enhanced. 
\begin{table}[t]
\centering
\begin{tabular}{c@{\hskip 5pt}c|c@{\hskip 5pt}c|c@{\hskip 5pt}c}
\hline
Parameter & Value & Parameter & Value & Parameter & Value
\\
\hline
$|V_{ud}|$ & 0.97370~\cite{PDG:2020}
&
$|V_{us}|$ & 0.2245~\cite{PDG:2020}
&
$|V_{cb}|$ & 0.0397\ts(6)~\cite{Iguro:2020cpg}
\\
\hline
$f_{D}$ & 0.2127~\cite{Bazavov:2017lyh}
&
$f_{\pi}$ & 0.1302\ts(8)~\cite{Aoki:2019cca}
&
$f_{K}$ & 0.1557\ts(3)~\cite{Aoki:2019cca}
\\
$f_{D_s}$ & 0.2499~\cite{Bazavov:2017lyh}  
&
$f_{\rho}$ & 0.213\ts(5)~\cite{Straub:2015ica}
&
$f_{K^*}$ & 0.204\ts(7)~\cite{Straub:2015ica}
\\
$f_{D^*}$ & 0.249~\cite{Gubernari:2018wyi}
&
$f_{\phi}$ & 0.233\ts(4)~\cite{Straub:2015ica}
&
$f_{\omega}$ & 0.197\ts(8)~\cite{Straub:2015ica}
\\
$f_{D_s^*}$ & 0.293~\cite{Gelhausen:2013wia}  
&&
\\
\hline
$F_0^{B D}(m_{\pi}^2)$ & 0.669\ts(10)~\cite{Iguro:2020cpg}
&
$F_0^{B D}(m_{K}^2)$ & 0.672\ts(10)~\cite{Iguro:2020cpg}
&
$F_1^{B D}(m_{\rho}^2)$ & 0.686\ts(10)~\cite{Iguro:2020cpg}
\\
$F_1^{B D}(m_{K^*}^2)$ & 0.692\ts(10)~\cite{Iguro:2020cpg}
&&
\\
$A_0^{B D^*}(m_{\pi}^2)$ & 0.725\ts(14)~\cite{Iguro:2020cpg}
&
$A_0^{B D^*}(m_{K}^2)$ & 0.732\ts(14)~\cite{Iguro:2020cpg}
&
$A_0^{B D^*}(m_{\eta}^2)$ & 0.734\ts(14)~\cite{Iguro:2020cpg}
\\
$A_0^{B D^*}(m_{\eta'}^2)$ & 0.754\ts(14)~\cite{Iguro:2020cpg}
&&
\\
\hline
$F_0^{B\pi}(m_{D}^2)$ & 0.288~\cite{Ball:2004ye}
&
$F_0^{B K}(m_{D}^2)$ & 0.364~\cite{Ball:2004ye}
&
$F_0^{B\eta}(m_{D}^2)$ & 0.192~\cite{Duplancic:2015zna}
\\
$F_0^{B\eta'}(m_{D}^2)$ & 0.148~\cite{Duplancic:2015zna}
&
$F_0^{B_s K}(m_{D}^2)$ & 0.310~\cite{Ball:2004ye,Khodjamirian:2017fxg}
&
$F_0^{B_s\eta}(m_{D}^2)$ & $-0.243$~\cite{Duplancic:2015zna}
\\
$F_0^{B_s\eta'}(m_{D}^2)$ & 0.290~\cite{Duplancic:2015zna}
&&
\\
$F_1^{B\pi}(m_{D^*}^2)$ & 0.328~\cite{Ball:2004ye}
&
$F_1^{B K}(m_{D^*}^2)$ & 0.420~\cite{Ball:2004ye}
&
$F_1^{B\eta}(m_{D^*}^2)$ & 0.210~\cite{Duplancic:2015zna}
\\
$F_1^{B\eta'}(m_{D^*}^2)$ & 0.162~\cite{Duplancic:2015zna}
&
$F_1^{B_s K}(m_{D^*}^2)$ & 0.357~\cite{Ball:2004ye,Khodjamirian:2017fxg}      
&
$F_1^{B_s\eta}(m_{D^*}^2)$ & $-0.264$~\cite{Duplancic:2015zna}
\\
$F_1^{B_s\eta'}(m_{D^*}^2)$ & 0.311~\cite{Duplancic:2015zna}
&&
\\
\hline
$A_0^{B\rho}(m_{D}^2)$ & 0.432~\cite{Straub:2015ica}
&
$A_0^{B\omega}(m_{D}^2)$ & 0.399~\cite{Straub:2015ica}
&
$A_0^{B K^*}(m_{D}^2)$ & 0.458~\cite{Straub:2015ica}
\\
$A_0^{B_s K^*}(m_{D}^2)$ & 0.438~\cite{Straub:2015ica}
&
$A_0^{B_s\phi}(m_{D}^2)$ & 0.515~\cite{Straub:2015ica}
\\
\hline
\end{tabular}
\caption{%
  The CKM matrix elements, 
  decay constants in units of GeV, 
  $\overline{B}\to D^{(*)}$ form factors, 
  $\overline{B}_{(s)}\to P$ form factors, and 
  $\overline{B}_{(s)}\to V$ form factors. 
  The definitions of the
  $\overline{B}_{(s)}\to\eta^{(\prime)}$ form factors include the mixing angle and the Clebsch-Gordan coefficients.
  We assume that the values of the $\overline{B}_s\to D_s^{(*)}$ form factors are identical to those of the
  $\overline{B}\to D^{(*)}$ form factors~\cite{Kobach:2019kfb,Bordone:2019guc}. 
}
\label{tab:parameters}
\end{table}

\subsection{Color-allowed decays}
\label{Sec:T}

As pointed out in ref.~\cite{Bordone:2020gao}, the theoretical predictions for the branching ratios of the color-allowed $\overline{B}\to DP$ and $\overline{B}\to D^*P$ decays are universally larger than the measured values. 
Let us first update the predictions in ref.~\cite{Bordone:2020gao} by adopting the latest results of the $\overline{B}\to D^{(*)}$ form factors given in ref.~\cite{Iguro:2020cpg}. 
The $\overline{B}\to D^{*}$ form factors in ref.~\cite{Iguro:2020cpg} have smaller uncertainties compared to those in ref.~\cite{Bordone:2019guc} because they include the full distribution data reported from Belle~\cite{Abdesselam:2017kjf,Waheed:2018djm} and QCD sum rule constraints on the parameters appearing in higher order corrections. 
It is assumed that the values of the $\overline{B}_s\to D_s^{(*)}$ form factors are identical to those of the $\overline{B}\to D^{(*)}$ form factors, because the $SU(3)$ breaking effects between them are found to be insignificant~\cite{Kobach:2019kfb,Bordone:2019guc}. 
Then, ignoring the rescattering contributions, 
the following predictions are obtained for 
the CP-averaged branching ratios of the color-allowed tree channels: 
\begin{align}
\mathcal{B}(\Bsbar \to D_s^+ \pi^-)
&=
\big( 40.9\pm 2.1 \big) \times 10^{-4},
\\
\mathcal{B}(\Bzerobar \to D^+ K^-)
&=
\big( 3.03\pm 0.15 \big) \times 10^{-4},
\\
\mathcal{B}(\Bsbar \to D_s^{*+} \pi^-)
&=
\big( 44.6\pm 2.2 \big) \times 10^{-4},
\\
\mathcal{B}(\Bzerobar \to D^{*+} K^-)
&=
\big( 3.27\pm 0.16 \big) \times 10^{-4},
\\
\mathcal{B}(\Bsbar \to D_s^+ \rho^-)
&=
\big( 104.5\pm 7.1 \big) \times 10^{-4},
\\
\mathcal{B}(\Bzerobar \to D^+ K^{*-})
&=
\big( 4.91\pm 0.41\big) \times 10^{-4}. 
\end{align}
The uncertainties come mainly from $|V_{cb}|$, $f_{P(V)}$ and the $B_{(s)}\to
D_{(s)}^{(*)}$ form factors in table~\ref{tab:parameters} as well as
from $a_1$ in eq.~\eqref{eq:a1univ}, where 
the correlations between $|V_{cb}|$ and the $B_{(s)}\to D_{(s)}^{(*)}$
form factors are included. 
The combined uncertainties are at the level of $5\%$ for $\overline{B}\to DP$ and $\overline{B}\to D^*P$ and
of $7$ to $8\%$ for $\overline{B}\to DV$. 
It is found that the predicted branching ratios for $\Bsbar \to D_s^+ \pi^-$ and $\Bzerobar \to D^+ K^-$ become slightly larger than the previous results in Ref.~\cite{Bordone:2020gao}.
Also the uncertainty in $\overline{B}\to D^*P$ is smaller due to the update of the form factors in Ref.~\cite{Iguro:2020cpg}.
Consequently, the above results deviate from the measured values by 
$3.5\ts\sigma$, $4.7\ts\sigma$, $4.5\ts\sigma$, $5.3\ts\sigma$, $2.3\ts\sigma$ and $0.5\ts\sigma$,
respectively. 
It is noted that the discrepancies are milder in the case of $\overline{B}\to DV$. 
Especially, the prediction for $\mathcal{B}(\overline{B}^0\to D^+ K^{*-})$ is consistent with the data. 

The above predictions are very clean theoretically. There is neither penguin nor annihilation contribution to these processes. As a result, there is no chirally-enhanced hard-scattering contributions at $\mathcal{O}(\Lambda_{\mathrm{QCD}}/m_B)$.
Moreover, power corrections at $\mathcal{O}(\Lambda_{\mathrm{QCD}}/m_B)$, including 
twist-3 two-particle contributions of light-meson light-cone distribution amplitudes, 
a hard-collinear gluon exchange between $b$ (or $c$) and the light meson, 
and a soft gluon exchange between the $B\to D$ system and the light meson, 
are expected to be less than a percent~\cite{Bordone:2020gao}. 
Besides, the QCD$\times$QED factorization is studied recently in ref.~\cite{Beneke:2021jhp}, where 
QED contributions to the color-allowed tree amplitudes are found to reduce the total amplitudes by the sub-percent level, though ultrasoft photons may correct the measured decay rates up to a few percent.

\subsection{Rescattering between color-allowed and color-suppressed decays}
\label{Sec:TC}

The rescattering effects modify the predictions for the color-allowed tree channels. 
In this subsection we investigate whether the quasi-elastic rescattering contributions based on the $SU(3)$ and $U(3)$ flavor symmetries can explain the discrepancies between the theoretical prediction and the experimental data discussed in the last subsection.

\begin{figure}[t]
\centering
\includegraphics[scale=0.55]{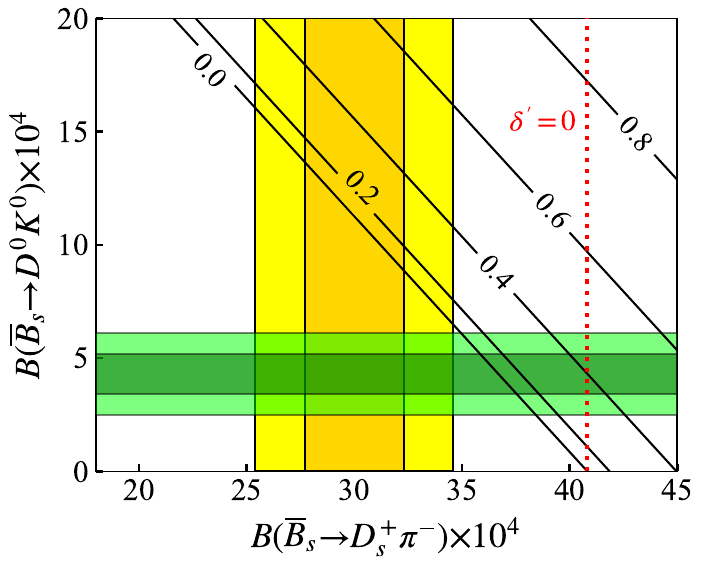}
\hfill
\includegraphics[scale=0.55]{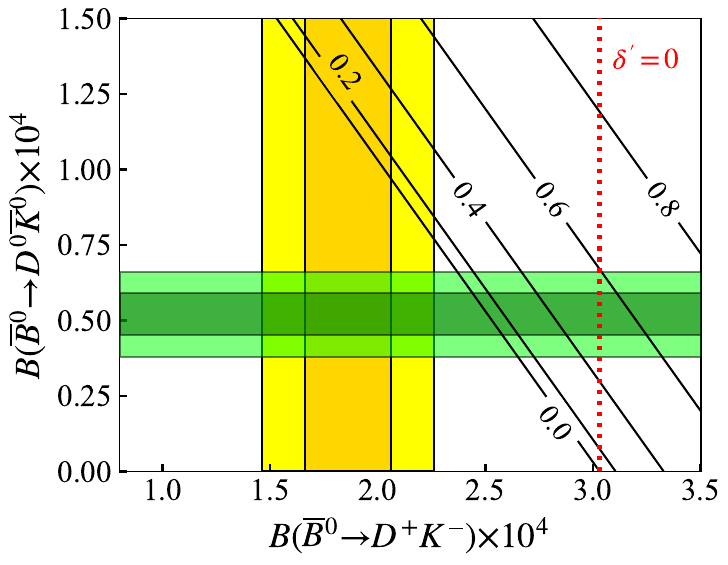}
\\[2mm]
\includegraphics[scale=0.55]{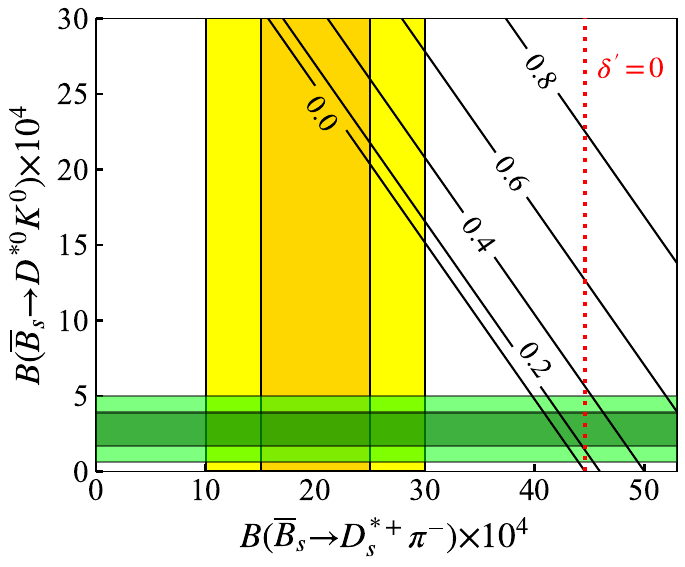}
\hfill
\includegraphics[scale=0.55]{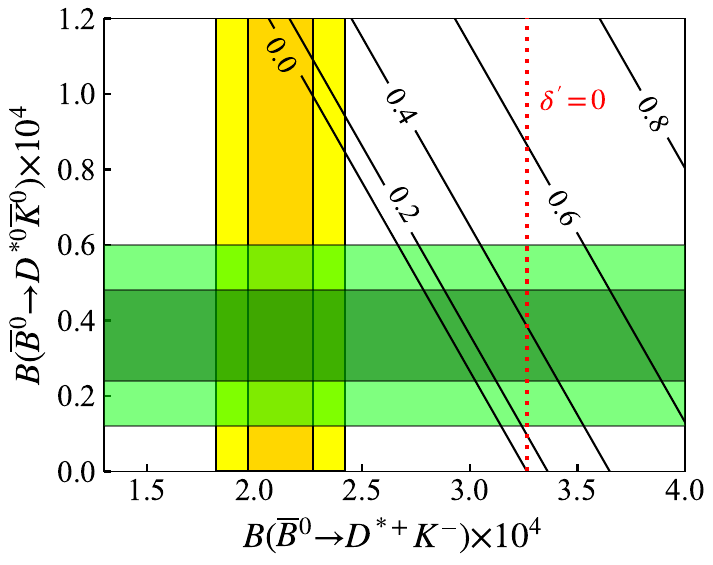}
\\[2mm]
\includegraphics[scale=0.55]{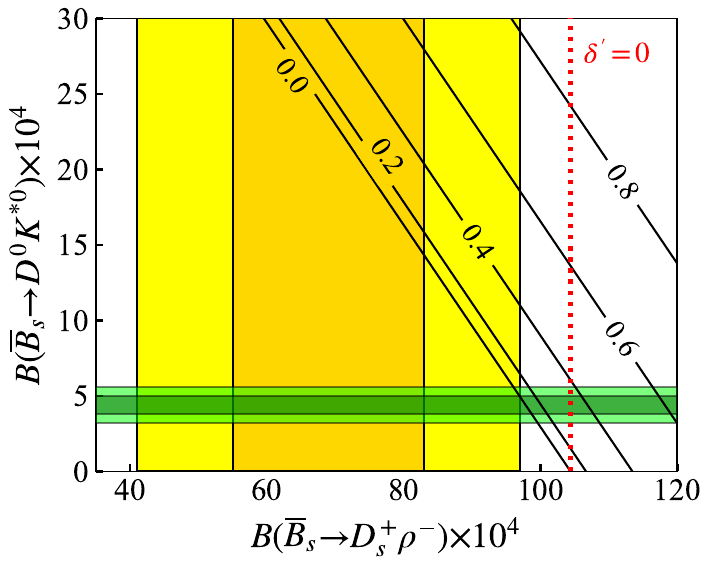}
\hfill
\includegraphics[scale=0.55]{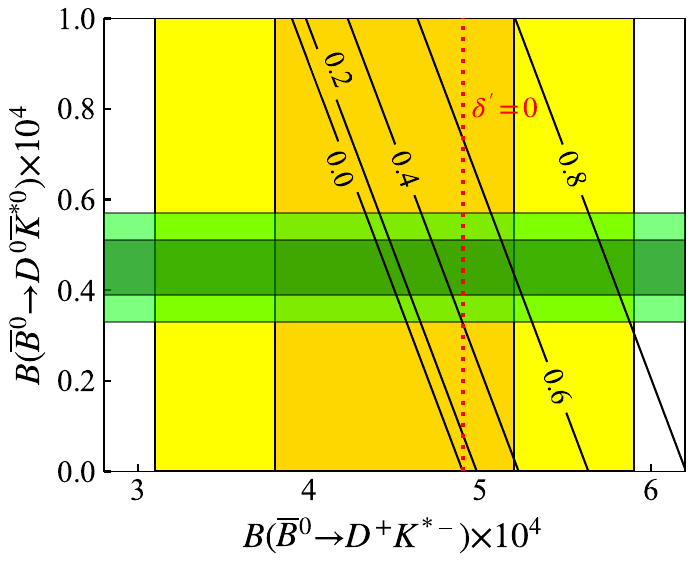}
\caption{Contours of the effective coefficient $|a_2^{\mathrm{eff}}|$ for the color-suppressed tree amplitude in the plane of the branching ratios for the color-allowed-tree and color-suppressed-tree channels, where the rescattering phase $\delta'$ and the relative phase between $a_1(m_b)$ and $a_2^{\mathrm{eff}}$ are taken to be free.
The vertical (yellow) and horizontal (green) bands show the experimental constraints on the branching ratios at the $1\ts\sigma$ (darker) and $2\ts\sigma$ (lighter) levels.
The vertical dotted line (red) corresponds to the case of no rescattering, i.e., $\delta'=0$.}
\label{fig:T-C-SM}
\end{figure}

As explained in section~\ref{Sec:res}, the color-allowed tree decay $\Bsbar \to D_s^+ \pi^-$ receives the quasi-elastic rescattering through the process $\Bsbar \to D^0 K^0 \to D_s^+ \pi^-$, where $\Bsbar \to D^0 K^0$ is a color-suppressed tree decay. 
The correlations between the color-allowed and the color-suppressed decays are shown in figure~\ref{fig:T-C-SM}.
Here the color-suppressed tree amplitude is parameterized with a single complex parameter $a_2^{\mathrm{eff}}$, which is substituted for $a_2(\mu)$ in eqs.~\eqref{eq:C_BDP}, \eqref{eq:C_BDstP} and \eqref{eq:C_BDV} and 
includes the factorizable and non-factorizable contributions. 
Then, $|a_2^{\mathrm{eff}}|$ and $\mathrm{arg}(a_2^{\mathrm{eff}})$ are taken as free parameters. 
Also, the rescattering contributions are parameterized by a single parameter $\delta'$ as in eq.~\eqref{eq:res2x2}.
In the numerical analyses, the CP-conserving strong phase $\mathrm{arg}(a_2^{\mathrm{eff}})$ and the rescattering phase $\delta'$ are scanned from $0$ to $2\pi$.
In particular, for $\overline{B}\to DV$, the solution 1 in eq.~\eqref{eq:U3sol1} is adopted because $\delta'=0$ is set in the solution 2. 
The uncertainties of the input parameters in table~\ref{tab:parameters} are ignored here. 

In figure~\ref{fig:T-C-SM}, the solid black lines denote the contours for
$|a_2^{\mathrm{eff}}|=0.0,\,0.2,\,0.4,\,0.6$ and $0.8$.
For example, in the case of $\Bsbar \to D_s^+ \pi^-$ and $\Bsbar \to D^0 K^0$, the branching ratios satisfy the following relation: 
\begin{align}
\frac
{
\mathcal{B}(\Bsbar \to D^0 K^0)
-
\mathcal{B}(\Bsbar \to D^0 K^0)\big|_{\delta'=0}
}
{
\mathcal{B}(\Bsbar \to D_s^+ \pi^-)
-
\mathcal{B}(\Bsbar \to D_s^+ \pi^-)\big|_{\delta'=0}
}
=
-
\frac
{p_{\mathrm{cm},D^0 K^0}}
{p_{\mathrm{cm},D_s^+ \pi^-}}
\left(\frac{f_D f_K}{f_{D_s} f_\pi}\right)^2,
\end{align}
where $p_{\mathrm{cm},D_s^+ \pi^-}$ and $p_{\mathrm{cm},D^0 K^0}$ are the magnitude of the three momentum of a final-state meson in the center-of-mass frame in each process.
If the rescattering contribution is absent, there are tensions between the theoretical predictions (denoted by a red dotted line) and the experimental data (denoted by yellow vertical bands) for the color-allowed tree decay.
Their branching ratios can be reduced by including the rescattering contribution, while the ratios for the color-suppressed channels increase at the same time. 
Although smaller $|a_2^{\mathrm{eff}}|$ is preferred from the data, even with $|a_2^{\mathrm{eff}}|=0.0$, there still remain a discrepancy between the theoretical results and the experimental data for $\overline{B}\to D^{(*)}P$.
Therefore, we conclude that the experimental data cannot be fully explained even if the quasi-elastic rescattering contributions are included.

The discrepancy may be explained by introducing rescattering contributions beyond the quasi-elastic approximation, large power corrections in the QCDF approach and/or NP contributions. 
Here the latter two cases are studied by regarding the coefficient $a_1(m_b)$ as an additional free parameter. 
\begin{figure}[t]
\centering
\includegraphics[scale=0.55]{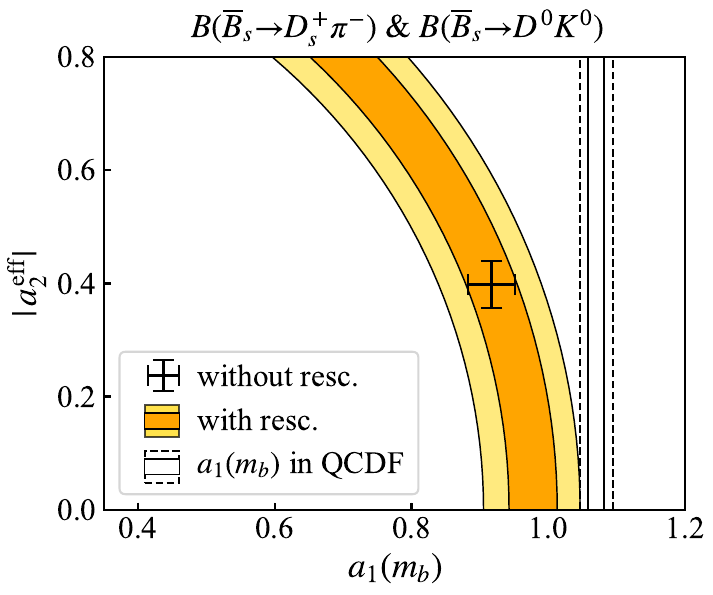}
\hfill
\includegraphics[scale=0.55]{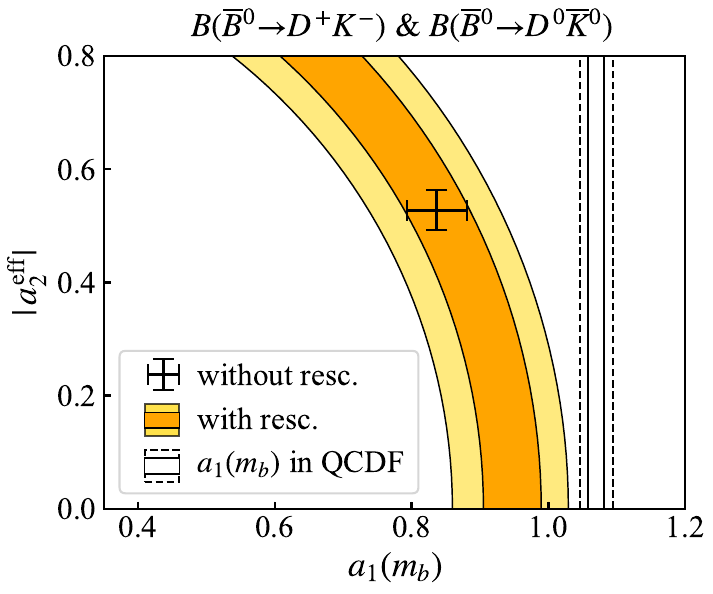}
\\[2mm]
\includegraphics[scale=0.55]{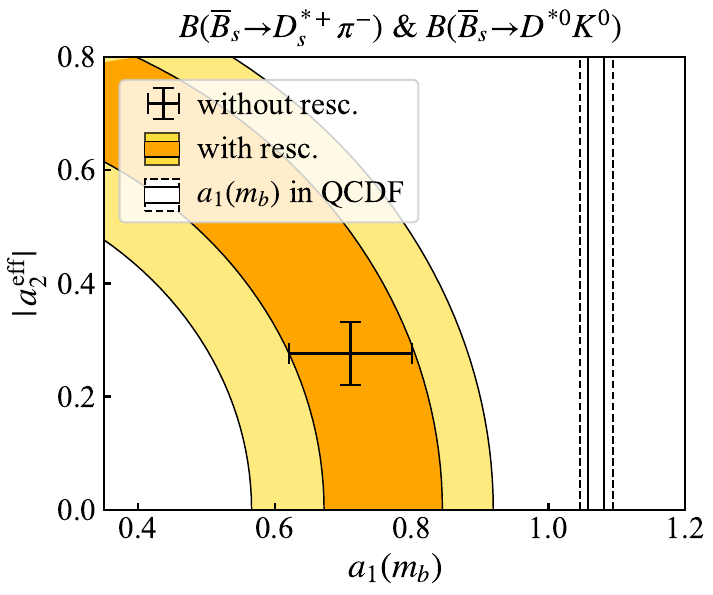}
\hfill
\includegraphics[scale=0.55]{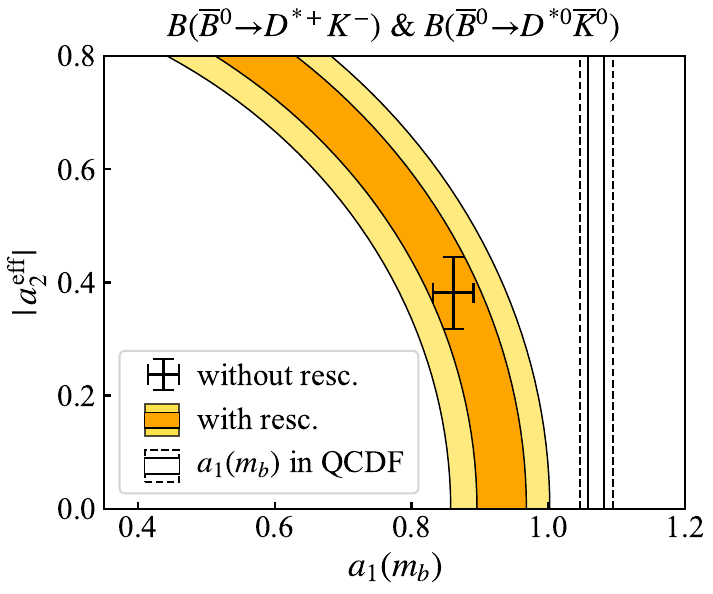}
\\[2mm]
\includegraphics[scale=0.55]{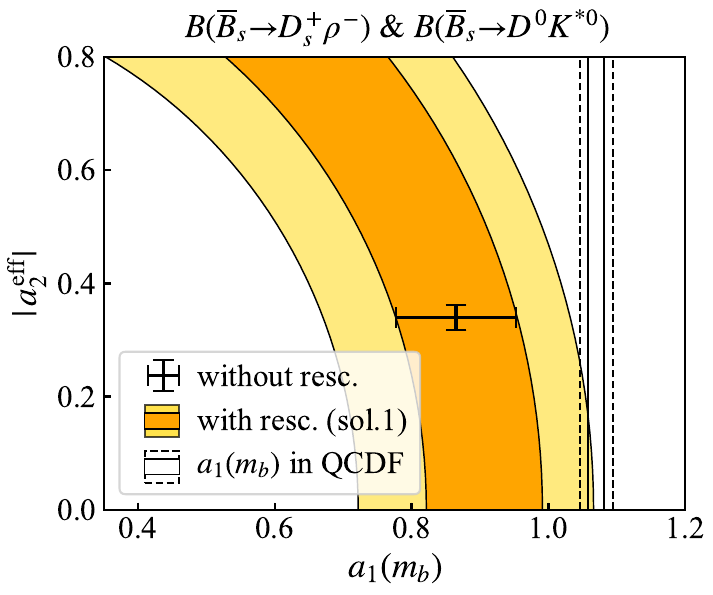}
\hfill
\includegraphics[scale=0.55]{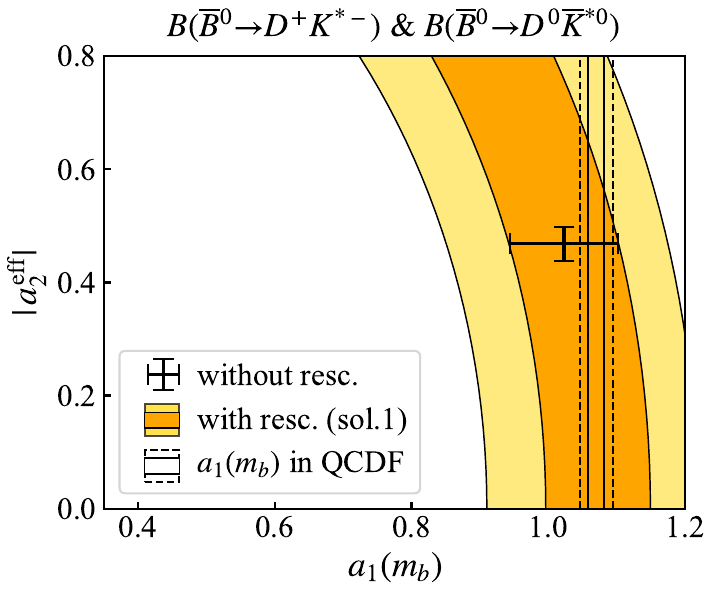}
\caption{The favored region in $a_1(m_b)$ vs $|a_2^{\mathrm{eff}}|$ plane considering color-allowed and color-suppressed channels simultaneously.
The orange regions show the experimental constraints at $1\ts\sigma$ (darker) and $2\ts\sigma$ (lighter) levels including the rescattering.
The cross in the band shows the result calculated without the rescattering. 
Black solid and dashed bands correspond to the QCDF prediction at the $1\ts\sigma$ and $2\ts\sigma$ levels.}
\label{fig:T-C-NP}
\end{figure}
In figure~\ref{fig:T-C-NP} the region of $a_1(m_b)$ and $|a_2^{\mathrm{eff}}|$ that accommodate the experimental data for the color-allowed and color-suppressed channels simultaneously is presented. 
For example, the colored regions in the top-left plot are calculated with the relation, 
\begin{align}
\alpha\ts r\ts
\big|a_1(m_b)\big|^2
+
\beta\ts \big|a_2^{\mathrm{eff}}\big|^2
=
r\,\mathcal{B}(\Bsbar \to D_s^+ \pi^-)
+
\mathcal{B}(\Bsbar \to D^0 K^0)\,,
\end{align}
where 
$r=(p_{\mathrm{cm},D^0 K^0}/p_{\mathrm{cm},D_s^+ \pi^-})
(f_D f_K/f_{D_s}/f_\pi)^2$, 
$\alpha = \mathcal{B}(\Bsbar \to D_s^+ \pi^-)\big|_{a_1(m_b)=1,\delta'=0}$ 
and $\beta = \mathcal{B}(\Bsbar \to D^0 K^0)\big|_{a_2^{\mathrm{eff}}=1,\delta'=0}$.
The orange regions show the experimental constraints on the branching ratios at the $1\ts\sigma$ (darker) and $2\ts\sigma$ (lighter) levels, where $\mathrm{arg}(a_2^{\mathrm{eff}})$ and $\delta'$ are scanned from $0$ to $2\pi$.  
Here and hereafter, new CP-violating phases are assumed to be absent for simplicity. 
It is observed that universal downward shifts of $\mathcal{O}(10\%)$ in $a_1(m_b)$ improve the overall consistency between the theoretical and experimental results. 
It is noted that the values of $|a_2^{\mathrm{eff}}|$ from each channel can be different from each other due to differences in the non-factorizable contributions among $\overline{B}\to DP$, $\overline{B}\to D^*P$ and $\overline{B}\to DV$ as well as to the $SU(3)$ breaking effects.

When the rescattering effects are neglected, one can determine $a_1(m_b)$ and $|a_2^{\mathrm{eff}}|$ individually from the branching ratios for the color-allowed and color-suppressed channels, respectively.
The result is denoted by the cross in the figure.
It is found that there are mild tensions among the values of $a_1(m_b)$ determined from each channel.
However, these tensions can be reduced by including the quasi-elastic rescattering.

\subsection{Global analysis}
\label{Sec:global}

As shown in the last subsection, the shift of $a_1(m_b)$ away from the QCDF prediction depends on the values of $|a_2^{\mathrm{eff}}|$, $\mathrm{arg}(a_2^{\mathrm{eff}})$ and $\delta'$.
The other decay channels listed in tables~\ref{tab:BDP}, \ref{tab:BDstP} and \ref{tab:BDV} provide constraints on these parameters under the $SU(3)$ and $U(3)$ flavor symmetries. In particular, the decays with the amplitude $T+C$, such as $B^- \to D^0 \pi^-$, give an information on correlations among $a_1(m_b)$, $|a_2^{\mathrm{eff}}|$ and $\mathrm{arg}(a_2^{\mathrm{eff}})$. 
It is noted that they do not receive rescattering contribution. 

In the analysis, the global-fitting package {\tt HEPfit}~\cite{deBlas:2019okz} is used with our own implementation of the $\overline{B}_{(s)}\to D^{(*)}_{(s)} M$ observables to investigate the constraints from the available data. 
The statistical analysis of the package is based on the {\tt BAT} library~\cite{Caldwell:2008fw}, which allows us to evaluate the posterior probability distributions of the parameters based on the Bayesian statistical inference with the Markov Chain Monte Carlo (MCMC).

The following two cases are considered: (i) a fit with only the six decay modes that have the amplitude $T$, $C$ or $T+C$, and (ii) a fit with all the modes for which the experimental data are available. 
The four parameters $a_1(m_b)$, Re$(a_2^{\mathrm{eff}})$, Im$(a_2^{\mathrm{eff}})$ and $\delta'$ are floated in the former case, while the nine parameters 
$a_1(m_b)$, Re$(a_2^{\mathrm{eff}})$, Im$(a_2^{\mathrm{eff}})$, Re$(E)$, Im$(E)$, $\delta'$, $\theta$, $\sigma$ and $\tau$ 
are floated in the latter. Flat priors are assumed for the fit parameters.
Other parameters are fixed to be constant in the fits.
The experimental constraints are listed in tables~\ref{tab:BDP}, \ref{tab:BDstP} and \ref{tab:BDV}.\footnote{A recent data $\mathcal{B}(B^-\to D^{*0}\pi^-)=(53.5\pm 2.2)\times 10^{-4}$ from LHCb~\cite{LHCb:2020hdx} is not included.} 
The data for $\mathcal{B}(\Bsbar \to D_s^{(*)+} K^-)$ are not included in the fits, because they are the sum of the branching ratios of $b\to c\bar{u}s$ and $b\to u\bar{c}s$ transitions. 
The branching ratio for $\Bsbar \to D_s^{(*)-} K^+$ is expected to be of the same order of magnitude as that for $\Bsbar \to D_s^{(*)+} K^-$. 
\begin{figure}[t]
\centering
\includegraphics[scale=0.55]{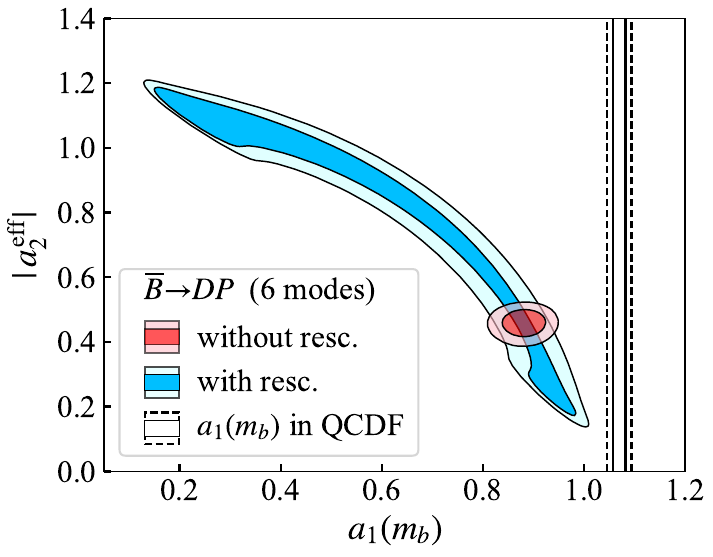}
\hfill
\includegraphics[scale=0.55]{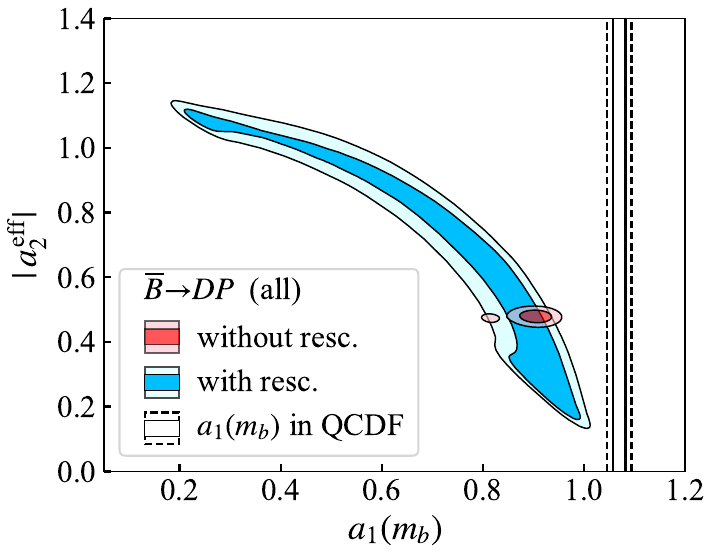}
\\[2mm]
\includegraphics[scale=0.55]{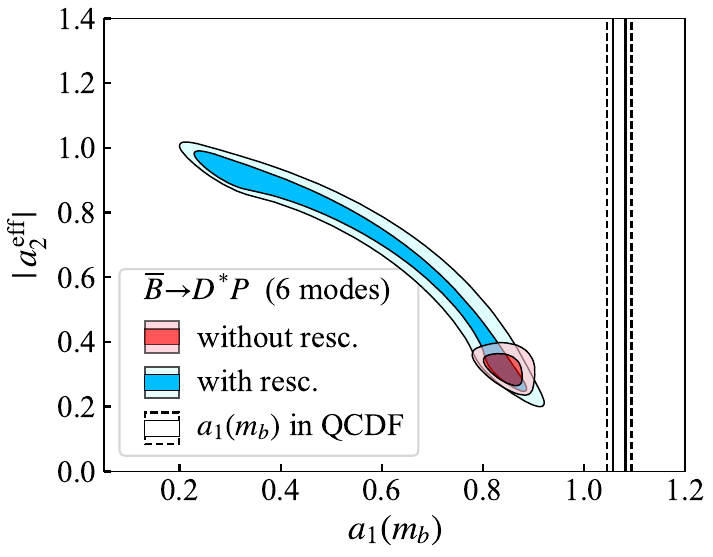}
\hfill
\includegraphics[scale=0.55]{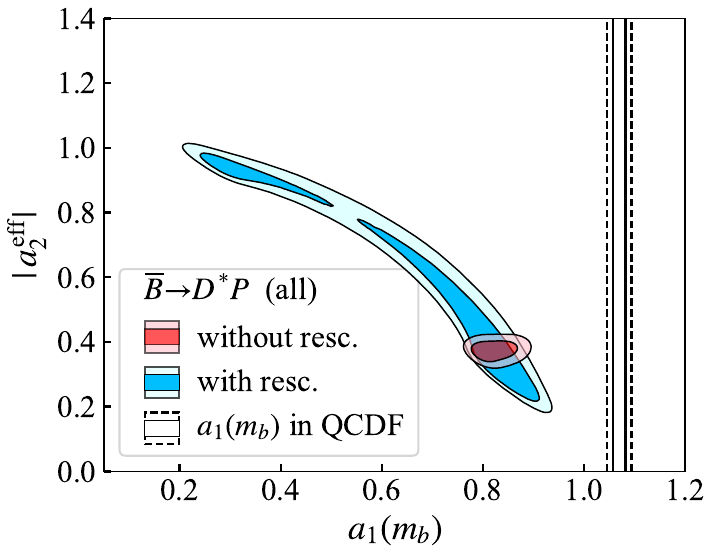}
\\[2mm]
\includegraphics[scale=0.55]{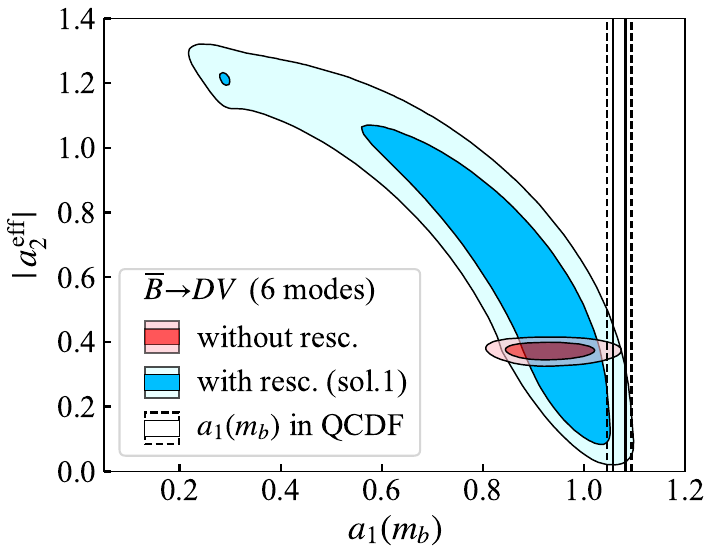}
\hfill
\includegraphics[scale=0.55]{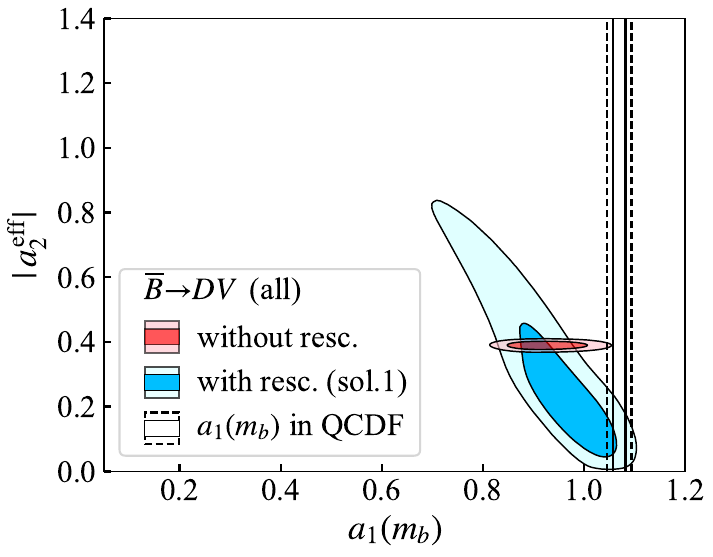}
\caption{Two-dimensional probability distributions of $a_1(m_b)$ and
  $|a_2^{\mathrm{eff}}|$. The darker (lighter) regions correspond to the 68\% (95\%) probability.
  For the rescattering parameters in $\overline{B}\to DV$, the solution 1 is adopted.}
\label{fig:NPfits}
\end{figure}
In figure~\ref{fig:NPfits} the two-dimensional probability distributions of $a_1(m_b)$ and $|a_2^{\mathrm{eff}}|$ are presented for the fit (i) in the left plots and for the fit (ii) in the right plots. 
The solution 1 in eq.~\eqref{eq:U3sol1} is adopted for $\overline{B}\to DV$. 
The results for the solution 2 in eq.~\eqref{eq:U3sol2}, which are not presented in figure~\ref{fig:NPfits}, become 
similar to those for the solution 1 without the rescattering, since the rescattering parameter $\delta'$ is fixed to zero. 
In each plot, the blue (red) region shows the fit result with (without) the rescattering.  
Here and in the following figures the darker (lighter) region corresponds to the 68\% (95\%) probability. 
By comparing the left plots in figure~\ref{fig:NPfits} with the plots in figure~\ref{fig:T-C-NP}, it is found that the preferred regions are diminished partly by including the $T+C$ modes. 
The regions in the right plots are basically similar to those in the left plots except for those of $\overline{B}\to DV$.
Thus, the fit is not improved even if the decay modes with the exchange amplitude $E$ are included, because the additional five fit parameters Re$(E)$, Im$(E)$, $\theta$, $\sigma$ and $\tau$ are introduced in the fit of $\overline{B}\to DP$ and $D^*P$, while the experimental data are not precise enough to constrain those parameters. 
The full one- and two-dimensional probability distributions obtained from the fits are summarized in appendix~\ref{Sec:fitresults}.
In the fit results of $\overline{B}\to DP$ and $D^*P$, 
it is observed that there are significant tensions between the theoretical predictions and the experimental data even with including the possible quasi-elastic rescattering effects. 

\begin{figure}[t]
\centering
\includegraphics[scale=0.6]{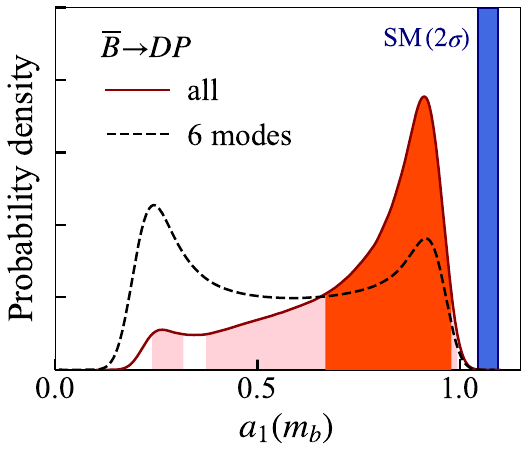}
\hspace{20mm}
\includegraphics[scale=0.6]{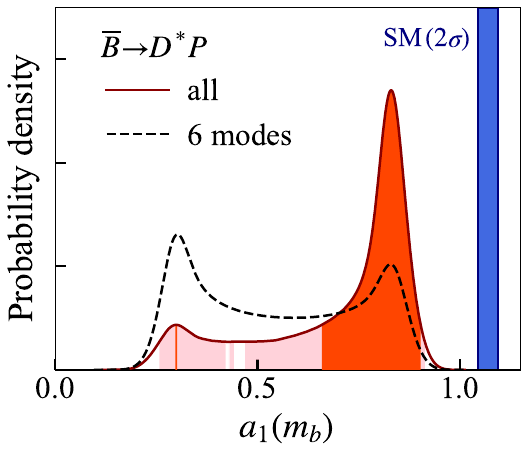}
\\[1mm]
\includegraphics[scale=0.6]{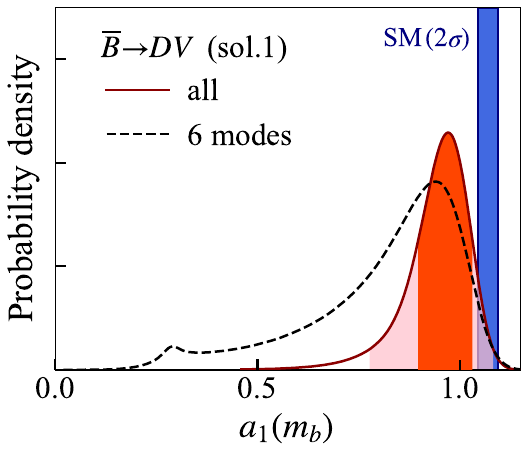}
\hspace{20mm}
\includegraphics[scale=0.6]{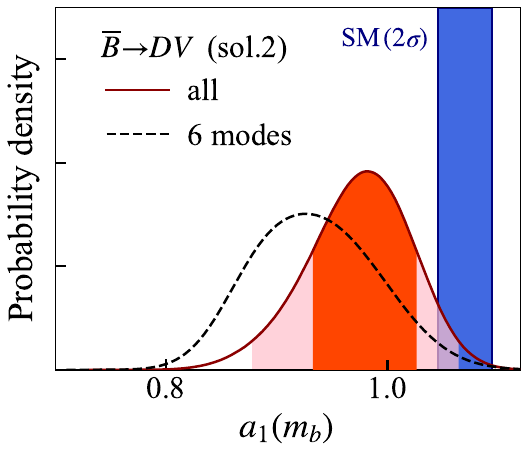}
\caption{Probability distributions of $a_1(m_b)$. The darker (lighter) regions correspond to the 68\% (95\%) probability. The blue vertical band represents the QCDF prediction of $a_1(m_b)$ at the $2\ts\sigma$ level.}
\label{fig:a1}
\end{figure}
The marginalized one-dimensional probability distributions of $a_1(m_b)$ in the fits (i) and (ii) are presented in figure~\ref{fig:a1}. 
Here the result for the solution 2 of $\overline{B}\to DV$ is also presented. 
In each plot the black dotted line shows the result for the fit (i). 
The distributions for $\overline{B}\to DP$ and $D^*P$ have also a peak in the low $a_1(m_b)$ region, where $|a_2^{\mathrm{eff}}|$ is larger than $a_1(m_b)$ as shown in figure~\ref{fig:NPfits}. 
The probability in the low $a_1(m_b)$ region is suppressed in the fit (ii) as presented by the red-colored distribution. 
It is found that the preferred regions of $a_1(m_b)$ differs from the QCDF prediction by more than $2\ts\sigma$ in $\overline{B}\to DP$ and $D^*P$, and $1\ts\sigma$ in $\overline{B}\to DV$.

The NP interpretations of the deviation in $a_1(m_b)$ are scrutinized in refs.~\cite{Bordone:2020gao,Iguro:2020ndk,Cai:2021mlt,Bordone:2021cca}.
The NP contributions of $\mathcal{O}(10\%)$ to the Wilson coefficients $C_1$ and $C_2$ have not been excluded by the current measurements of other observables~\cite{Bobeth:2014rda,Brod:2014bfa,Lenz:2019lvd}.
Numerically, the NP contribution to $a_1(m_b)$ is given in terms of the coefficients at the NP scale, which is taken to be 1~TeV as a reference~\cite{Iguro:2020ndk}: 
\begin{align}
  a_1^{\mathrm{NP}}(m_b)=&\,
  -0.19\, C_1^{\mathrm{NP}}(1\,\mathrm{TeV})
  +1.07\, C_2^{\mathrm{NP}}(1\,\mathrm{TeV})\,,
\end{align}
where the renormalization-group evolution is calculated in the leading-logarithmic approximation.
In order to achieve $a_1^{\mathrm{NP}}(m_b)\sim -0.1$, it is estimated that $C_2^{\mathrm{NP}}(1\,\mathrm{TeV})\sim -0.1$ is necessary, which is about 10\% of the SM value. 
Since the SM contribution to $C_2$ is generated through a tree-level $W$-boson exchange, the requested NP particles must have strong couplings to the light quarks. 
Such scenarios confront with constraints from LHC, meson mixings and/or $K$ and also other $B$ decays~\cite{Brod:2014bfa,Iguro:2020ndk}, and thus, it is challenging in general to construct a viable NP scenario.

\section{Summary}
\label{Sec:summary}

Recent developments in the study of the $\overline{B}_{(s)}\to D_{(s)}$ transition form factors have shed light on the discrepancies between the theoretical predictions and the experimental data for the color-allowed tree decays $B(\overline{B}^0 \to D^{(*)+}K^-)$ and $B(\overline{B}^0_s \to D^{(*)+}_s \pi^-)$. 
The theoretical predictions for these branching ratios are universally larger than the measured values. 
On the other hand, it has been known that the decays with the color-suppressed tree amplitude have smaller branching ratios in the factorization approaches than the experimental data. 
This motivated us to revisit the final-state rescattering effects in the $\overline{B}\to DP$, $\overline{B}\to D^*P$ and $\overline{B}\to DV$ channels.

We have examined whether the discrepancies can be explained by the quasi-elastic rescattering contributions based on the $SU(3)$ and $U(3)$ flavor symmetries. 
From the analysis of the color-allowed modes 
$\overline{B}^0 \to D^{(*)+}K^-$ and 
$\overline{B}^0_s \to D^{(*)+}_s \pi^-$
and the color-suppressed modes 
$\overline{B}^0 \to D^{(*)0}\overline{K}^0$ and 
$\overline{B}^0_s \to D^{(*)0}K^0$, 
we have concluded that the rescattering effects cannot account for the discrepancies. 
Hence there should exist additional contributions such as inelastic rescattering contributions beyond the quasi-elastic picture, large power corrections in the QCDF approach and/or NP contributions to the short-distance Wilson coefficients. 
We have studied the latter two possibilities by taking the coefficient $a_1(m_b)$ as a fit parameter. 
The preferred range of $a_1(m_b)$ has been determined from the fits to all the available data for $\overline{B}_{(s)}\to D^{(*)}_{(s)} M$, where the fits have been performed separately for $\overline{B}\to DP$, $\overline{B}\to D^*P$ and $\overline{B}\to DV$.
The results point to a downward shift of $\mathcal{O}(10\%)$ in $a_1(m_b)$ compared to the QCDF prediction.
Remarkably, all the fits are consistent with each other.

As argued in detail in ref.~\cite{Bordone:2020gao}, it is hard to resolve the discrepancies by (unknown) higher-order or higher-power corrections to the decay amplitudes in the QCDF approach. 
Effects of the inelastic rescattering might be significant, though their discussion is beyond the scope of this paper. 
On the other hand, it is also challenging to construct viable NP models that realizes the NP contribution of $\mathcal{O}(10\%)$. 
Therefore, further theoretical and experimental studies are needed to clarify this issue.

\section*{Acknowledgements}

We are grateful to Marzia Bordone, Nico Gubernari, 
Martin Jung and Danny van Dyk 
for helpful information on 
their calculation of the $\overline{B}_{(s)}\to D_{(s)}$ form factors. 
We also thank Teppei Kitahara for fruitful discussion.
This work is supported in part by the Japan Society for the Promotion of Science
under 
the Grant-in-Aid for Scientific Research on Innovative Areas (No.~21H00086 [ME]),
the Grant-in-Aid for Scientific Research B (No.~21H01086 [ME]), 
the Grant-in-Aid for Early-Career Scientists (No.~16K17681 [ME]), 
Research Fellowships for Young Scientists (No.~19J10980 [SI]), 
Core-to-Core Program (No.~JPJSCCA20200002 [SI]), 
the World Premier International Research Center Initiative [SI],
and 
the Grant-in-Aid for Scientific Research C (No.~17K05429 [SM]). 
SI would like to thank the warm hospitality at the KEK theory center, 
where he stayed during the initial stage of this project.

\appendix
\section{Fit results}
\label{Sec:fitresults}

In this appendix we present some details of the fit results in section~\ref{Sec:global}.

\begin{figure}[t]
\centering
\includegraphics[scale=0.88]{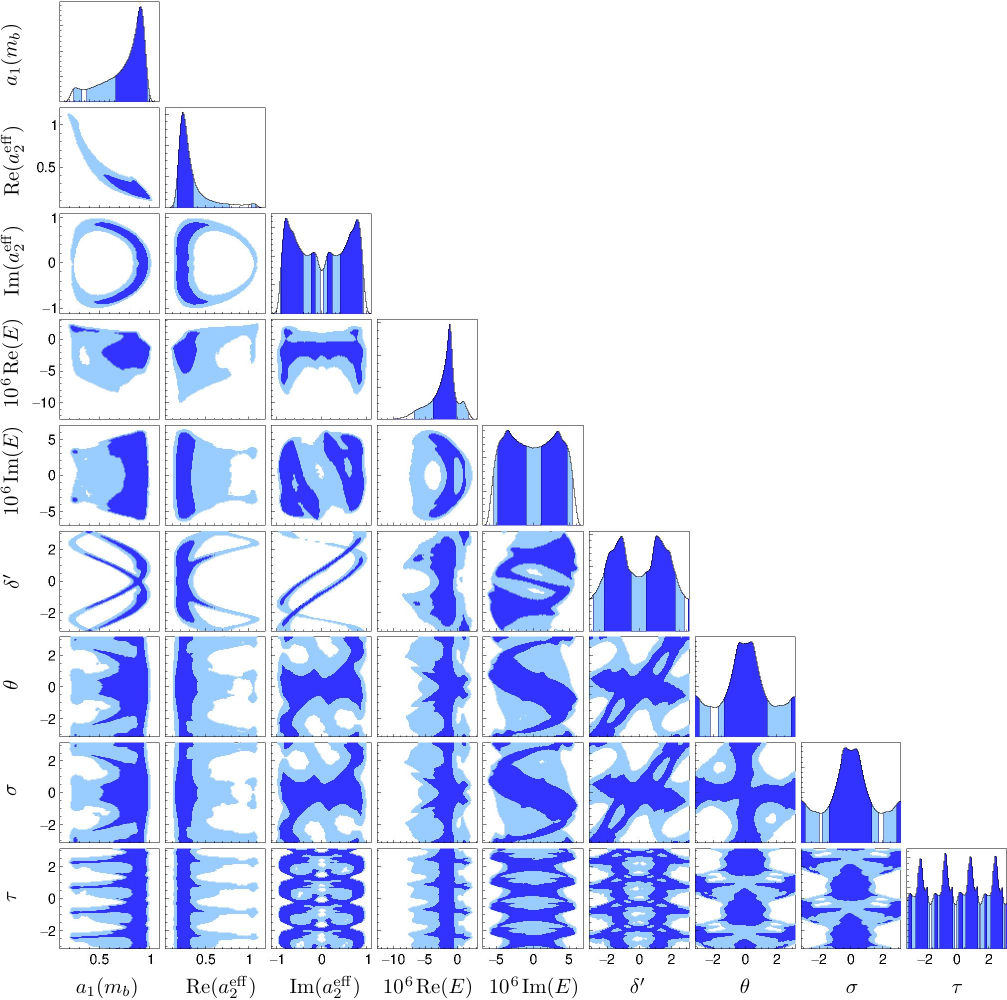}
\caption{Probability distributions of the parameters in $\overline{B}\to DP$. The darker (lighter) regions correspond to the 68\% (95\%) probability.}
\label{fig:all-DP}
\end{figure}

\begin{figure}[t]
\centering
\includegraphics[scale=0.88]{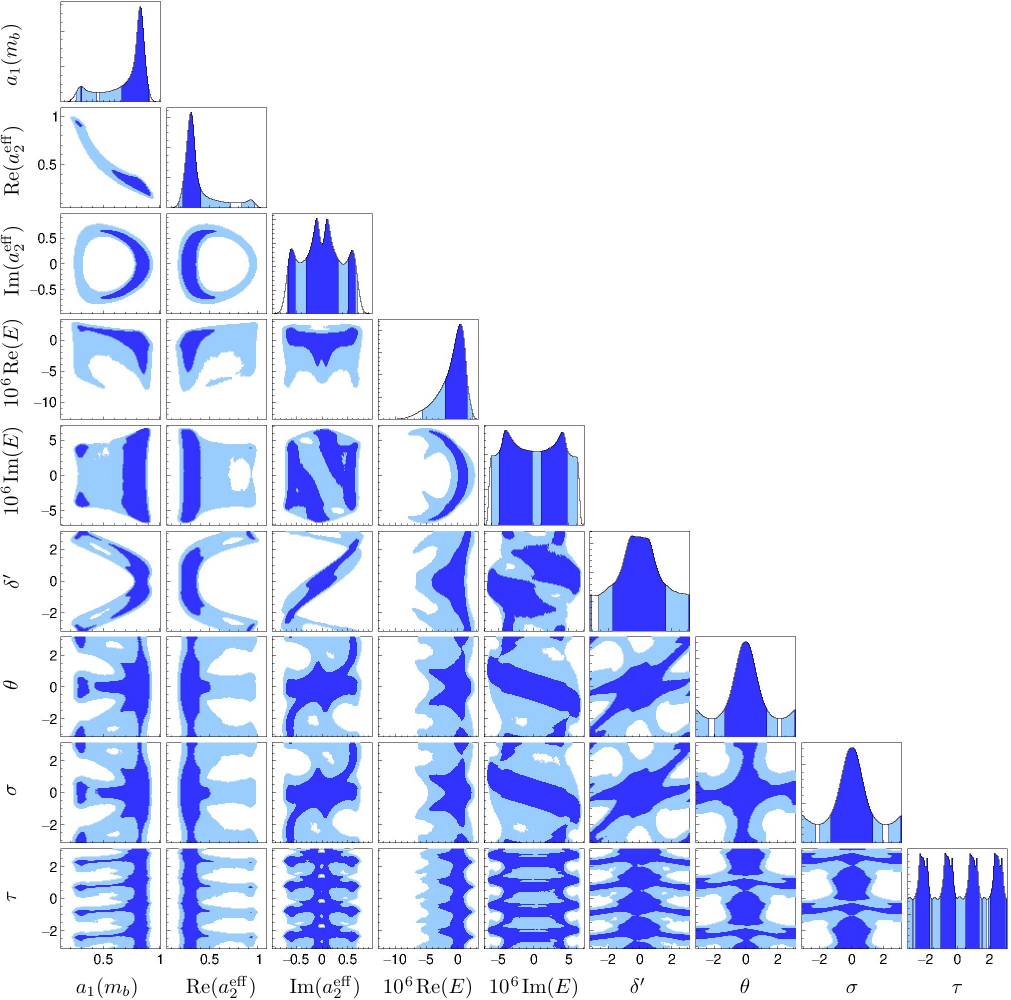}
\caption{Same as figure~\ref{fig:all-DP}, but for $\overline{B}\to D^*P$.}
\label{fig:all-DStarP}
\end{figure}

\begin{figure}[t]
\centering
\includegraphics[scale=1.0]{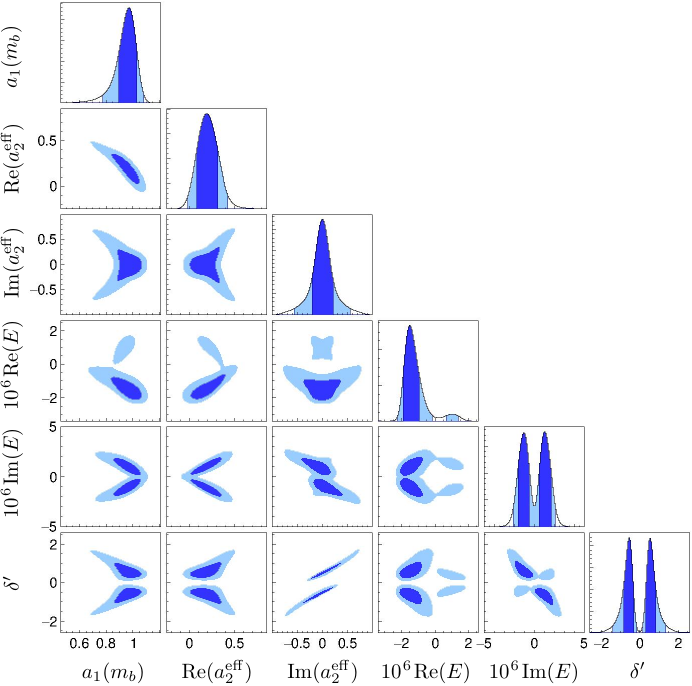}
\caption{Same as figure~\ref{fig:all-DP}, but for $\overline{B}\to DV$ with the solution 1 in eq.~\eqref{eq:U3sol1}.}
\label{fig:all-DV-sol1}
\end{figure}

\begin{figure}[t]
\centering
\includegraphics[scale=1.0]{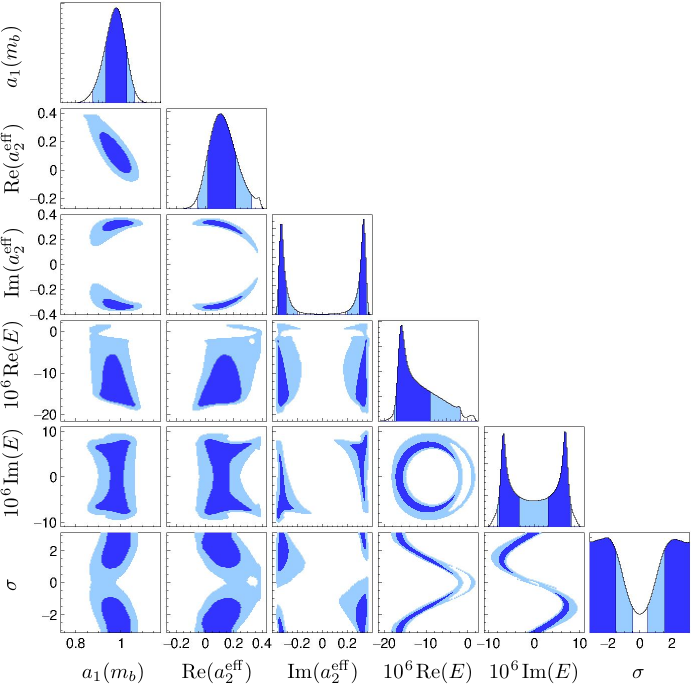}
\caption{Same as figure~\ref{fig:all-DP}, but for $\overline{B}\to DV$ with the solution 2 in eq.~\eqref{eq:U3sol2}.}
\label{fig:all-DV-sol2}
\end{figure}

Figures~\ref{fig:all-DP}, \ref{fig:all-DStarP}, \ref{fig:all-DV-sol1}, and \ref{fig:all-DV-sol2} show the one-dimensional and two-dimensional probability distributions of the parameters in the fit (ii), which uses all the available experimental data for $\overline{B}\to DP$, $\overline{B}\to D^*P$, $\overline{B}\to DV$ with the solution 1 in eq.~\eqref{eq:U3sol1}, and $\overline{B}\to DV$ with the solution 2 in eq.~\eqref{eq:U3sol2}, respectively. 
In each plot the darker (lighter) regions correspond to the smallest region of the 68\% (95\%) probability. 
There is only a single rescattering parameter $\delta'$ or $\sigma$ in the fits to the $\overline{B}\to DV$ data, because the $U(3)$ flavor symmetry is adopted instead of $SU(3)$. 
The results for $\overline{B}\to DP$ and $\overline{B}\to D^*P$ have a long tail in the one-dimensional distributions of $a_1(m_b)$, which corresponds to a large value of $a_2^{\mathrm{eff}}$. 

\begin{figure}[t]
\centering
\includegraphics[scale=0.54]{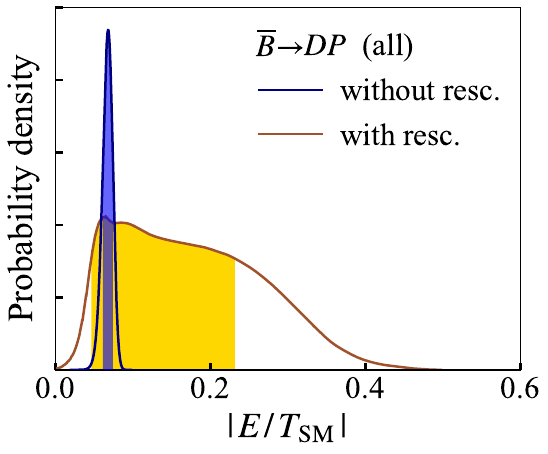}
\hfill
\includegraphics[scale=0.54]{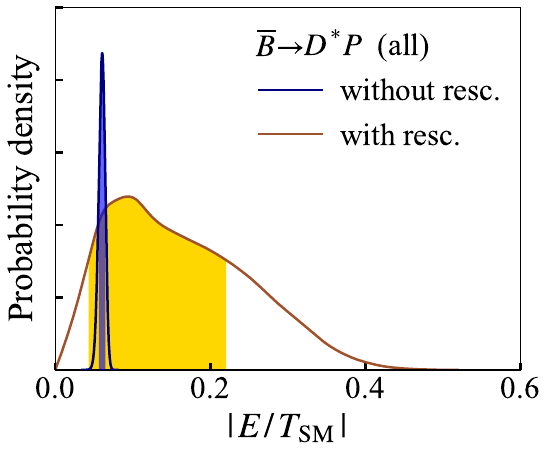}
\hfill
\includegraphics[scale=0.54]{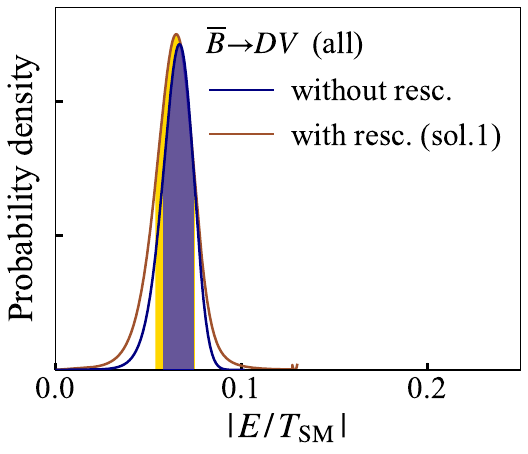}
\caption{Probability distributions of $|E/T_{\mathrm{SM}}|$.  
The colored regions correspond to 68\% probability.}
\label{fig:E}
\end{figure}

In addition, the one-dimensional distributions of $|E/T_{\mathrm{SM}}|$ are shown in figure~\ref{fig:E}, where $T_{\mathrm{SM}}$ is the SM value of the color-allowed tree amplitude for $\Bzerobar \to D^+ K^-$, $\Bzerobar \to D^{*+} K^-$ or $\Bzerobar \to D^+ K^{*-}$. 
Here the results in the fits with (without) the rescattering are presented as the distributions in yellow (blue). 
It is noted that the exchange amplitude $E$ is power-suppressed compared to the color-allowed tree amplitude~\cite{Beneke:2000ry}. 
The ranges of $|E/T_{\mathrm{SM}}|$ in the fits without the rescattering are consistent with the naive power-counting expectation, while a mildly larger value of $|E/T_{\mathrm{SM}}|$ is allowed in the fits with the rescattering.

Further experimental information on the $\overline{B}_s$ decays with the $b\to c\bar{u}s$ transition listed in tables~\ref{tab:BDP}, \ref{tab:BDstP} and \ref{tab:BDV} is helpful to obtain stronger constraints on the fit parameters, though it would be challenging to measure those decays at the LHCb experiment in most cases. 
Theoretical studies of the color-suppressed-tree and the exchange amplitudes would also be helpful to set stronger constraints. 

\bibliography{ref}
\bibliographystyle{JHEP}

\end{document}